\documentclass{article}
\usepackage[ruled]{algorithm2e}

\SetAlFnt{\small}
\SetAlCapFnt{\small}
\SetAlCapNameFnt{\small}
\SetAlCapHSkip{0pt}
\IncMargin{-\parindent}

\usepackage{amsthm}

\newtheorem{definition}{Definition}

\usepackage[latin1]{inputenc}
\usepackage{amsmath}
\usepackage{amssymb}
\usepackage{mathbbol}
\usepackage{tikz} 
\usepackage[scaled=1]{helvet}
\usepackage{verbatim}
\usepackage{subfig}

\newcommand{\myldots}{...}
\def\sep{}  
\def\allInd{\ast }
\def\lset{\mathcal L}
\def\lsetset{\mathbb{L}}
\def\eeins{\mathbf{I}}

\def\flatRBAC{\text{flat}}

\def\tdi{T\!DI}
\def\upa{U\!P\!A}
\def\ua{U\!A}
\def\pa{P\!A}
\def\user{U\mspace{-0.9mu}S\mspace{-0.9mu}E\mspace{-0.9mu}R\mspace{-0.7mu}S}
\def\rc{RC}
\def\perm{P\mspace{-0.9mu}R\mspace{-0.9mu}M\mspace{-0.9mu}S}
\def\roles{RO\mspace{-0.9mu}L\mspace{-0.9mu}E\mspace{-0.9mu}S}

\def\BMP{\ast } 
\newcommand{\expect}[1]{\mathbb E \left[ #1 \right]} 
\newcommand{\expectz}[2]{\mathbb{E}_{ #1 } \left[ #2 \right]} 

\begin{document}

\author{
Mario Frank$^{\sharp}$, Joachim M. Buhmann$^{\flat}$,
David Basin$^{\flat}$
\\
$^{\sharp}$ UC Berkeley, Computer Science Division \\ 
$^{\flat}$ETH Zürich, Department of Computer Science
}

\title{Role Mining with Probabilistic Models}

\maketitle

\begin{abstract}
     Role mining tackles the problem of finding a role-based access control
  (RBAC) configuration, given an access-control matrix assigning users
  to access permissions as input. 
  Most role mining approaches work by constructing a large set of
  candidate roles and use a greedy selection 
  strategy to iteratively pick
  a small subset such that the differences between the resulting RBAC
  configuration and the access control matrix are minimized.
  In this paper, we advocate an alternative approach that recasts role
  mining as an inference problem rather than a lossy compression
  problem. Instead of using combinatorial algorithms to minimize the
  number of roles needed to represent the access-control matrix, we
  derive probabilistic models to learn the RBAC configuration that
  most likely underlies the given matrix.

  Our models are generative in that they reflect the way that
  permissions are assigned to users in a given RBAC configuration. We
  additionally model how user-permission assignments that conflict
  with an RBAC configuration emerge and we investigate the influence of
  constraints on role hierarchies and on the number of assignments.
  In experiments with access-control matrices from real-world
  enterprises, we compare our proposed models with other role mining
  methods.  Our results show that 
  our probabilistic models infer roles that generalize well to new
  system users for a wide variety of data, while other models'
  generalization abilities depend on the dataset
  given. 
\end{abstract}

\section{Introduction}
Role-Based Access Control (RBAC) \cite{rbacOrig} is a popular access
control model.  Rather than directly assigning users to permissions
for using resources, {for example} via an access control matrix, in RBAC one
introduces a set of roles.
The roles are used to decompose a user-permission relation
into two relations: a user-role relation that assigns users to roles
and a role-permission relation that assigns roles to permissions.
Since roles are (or should be) natural abstractions of functional
roles within an enterprise, these two relations are conceptually
easier to work with than a direct assignment of users to permissions.
Experience with RBAC indicates that this decomposition works well in
practice and facilitates the administration of large-scale
authorization policies for enterprises with many thousands of users
and permissions.

Although the benefits of using RBAC are widely recognized, its
adoption and administration can be problematic in practice.  Adoption
requires that an enterprise migrates its authorizations to RBAC.
Additionally, even after RBAC is in place, authorizations may need to
be reassigned after major changes within the enterprise, for example, after a 
reorganization or a merger where processes and IT systems from
different divisions must be consolidated.  Migration and the need to
reengineer roles after major changes are the 
{most expensive} aspects of RBAC.

To address these problems, different approaches have been developed to
configuring RBAC systems.  These approaches have been classified into
two kinds \cite{vaidya06roleminer}: top-down and bottom-up role
engineering.  Top-down engineering configures RBAC independently of
any existing user-permission assignments by analyzing the enterprise's
business processes and its security policies.  Bottom-up
role-engineering uses existing user-permission assignments to find a
suitable RBAC configuration.

When carried out manually, bottom-up role engineering is
very difficult and therefore incurs high costs and poses security risks. 
To simplify this step, \cite{kuhlmann} proposed the first automatic
method for bottom-up role engineering and coined the term \emph{role
  mining} for such methods. Numerous role mining algorithms have been
developed since then, notably
\cite{EneEtAl,JSVaidyaSacmat07rm,VaidyaBoolDecomp,Molloy:2010:MRN,VaidyaPrioSubset}.

Unfortunately, most role mining algorithms suffer from the drawback
that the discovered roles are artificial and unintuitive for the
administrators who must assign users to roles.  This
undesirable effect is linked to the design principle for role mining
algorithms that aim at achieving one of the following two goals:
\vspace{-0.5em}
\begin{itemize}
\item minimizing the deviation between the RBAC configuration and the
  given access control matrix for a given number of roles, or
\item minimizing the number of roles for a given deviation (possibly zero).
\end{itemize}
\vspace{-0.5em} Both goals amount to data compression.  As a result,
the roles found are often synthetic sets of permissions.  Although
these roles minimize the RBAC configuration, they are difficult to
interpret as functional roles, also called \emph{business roles}, that
reflect the business attributes of the users assigned to them. This
shortcoming not only limits the acceptance of the new RBAC
configuration among employees, it also makes the RBAC configuration difficult to
maintain, for example, to update when users join the system or change their
business role within the enterprise.  Moreover, by optimizing the
roles to fit existing authorizations as close as possible, existing
erroneous user-permission assignments are likely to be migrated to the
new RBAC system.  Several attempts have been made to incorporate
business attributes into the role mining process to improve the
business-relevance of the roles \cite{molloy,HyDRo}.
{However, these algorithms compress the access control matrix as the
underlying objective and they do not necessarily infer predictive
roles.}

{The problems outlined above} arise because the predominant definitions of the role
mining problem {\cite{Vaidya:2010:RMP:1805974.1805983,VaidyaHierarchy,molloyTechRep,Lu:2012:CRM:2360740.2360784}} do
not reflect realistic problem settings. In particular, all prior
definitions assume the existence of input that is usually not
available in {practice}.  For example, the input includes
either the number of roles to be found or the 
{maximally tolerable residual error}
when fitting the RBAC configuration to the given
access-control matrix.  
{Guided by these definitions, the respective algorithms focus on
  compressing user-permission assignments rather than on structured
  predictions of roles. However, compression addresses the wrong problems.}  We
argue that there is a need for a definition of role mining {as an
  inference problem} that is based on realistic assumptions.
Consequently, given such a definition, algorithms must be developed
that aim at solving this problem. Moreover it is necessary to have
quality measures to compare how well different algorithms solve the
problem.  Our contributions cover all these aspects:
\vspace{-0.5em}
\begin{itemize}
\item {We provide the first complete approach to probabilistic
  role mining, including the problem definition, models and algorithms,
  and quality measures.}
\item We define role mining as an inference problem: learning the roles that
  most likely {explain} 
  the given data. We carefully motivate and explicate the assumptions
  involved.
\item We propose probabilistic models for finding the RBAC
  configuration that {has}  most likely {generated} 
  a given access-control matrix.
\item We {demonstrate} that the RBAC configuration that best solves the role
  inference problem {contains} roles that generalize well to a set of
  hold-out users. We therefore provide a generalization test to
  evaluate role mining algorithms and apply this test to our methods
  and to competing approaches.
\item {We experimentally demonstrate that our probabilistic
    approach provides a sound confidence estimate for given
    user-permission assignments.}
\item We develop a hybrid role mining algorithm that incorporates
  business attributes into the role mining process and thereby
  improves the interpretability of roles.
\end{itemize}

We proceed as follows. First, in Section~\ref{rmDef_newDefmain}, we review the assumptions that are often
implicitly made for role mining and {we} use them to define the
inference role mining problem together with appropriate evaluation
criteria. Then, in Section~\ref{sec_probmodels}, we derive a class of
probabilistic models from the deterministic user-permission assignment
rule of RBAC.  We explain two model instances in detail and we propose
learning algorithms for them in Section~\ref{sec_optimization}.  In
Sections~\ref{sec_expsFull}, we report on experimental findings on
real world access control data.  In Section~\ref{hybrid_sec:risk}, we
show how to include business attributes of the users in the role
mining process to obtain a hybrid role mining algorithm and we
experimentally investigate how these attributes influence the RBAC
configurations discovered.  We discuss related work in
Section~\ref{relwork} and draw conclusions in Section~\ref{concl-sec}

\section{Problem definition}
\label{rmDef_newDefmain}
In this section we define the role inference problem.  First, we
explain the assumptions underlying our definition.  Afterwards we
present a
general definition of role mining that takes
business-relevant information into account. The pure bottom-up problem
without such information will be a special case of the general problem.
Finally, we propose quality measures for assessing role mining algorithms.

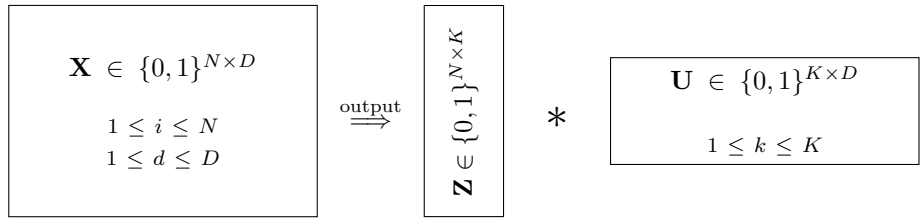
\begin{figure}[t]
  \centering
   \tikzstyle{xblock} = [rectangle, draw, text width=11em, text centered, minimum height=8em]
   \tikzstyle{zblock} = [rectangle, draw, text width= 7.325em, text centered, minimum height=3em]
   \tikzstyle{ublock} = [rectangle, draw, text width=11em, text centered, minimum height=3em]
  \tikzstyle{line} = [draw, thick, ->]
  \begin{tikzpicture}[node distance = 4cm, auto]
    \node [xblock] (input) {$\mathbf{X}\in \{0,1\}^{N\times D}$
    \\ \ \\{{\footnotesize $1\!\!\leq\!\! i\!\!\leq\!\! N$ \\ $1\!\!\leq\!\! d\!\!\leq \!\!D$}}
};
    \node [zblock, right of=input, rotate=90] (zMat) {\mbox{$\mathbf{Z}\in
\{0,1\}^{N\times K}$}};
    \node [ublock, right of=zMat] (uMat) {$\mathbf{U}\in\{0,1\}^{K\times D}$
     \\ \ \\{{\footnotesize $1\!\!\leq\!\! k\!\!\leq\!\! K$}}
};
    \path (input) -- node [above=-10pt] { $\stackrel{\textrm{output}}{\Longrightarrow}$ } (zMat);
    \path (zMat)  -- node [above=-10pt] {\LARGE{$\BMP$}} (uMat);
  \end{tikzpicture}
  \caption{Graphical illustration of the decomposition of an access
    control matrix $\mathbf{X}$, which assigns $N$ users to $D$ permissions, into an RBAC configuration with two assignments: $\mathbf{Z}$ assigns users to $K$ roles and  $\mathbf{U}$ assigns roles to permissions. The reconstruction $\mathbf{Z} \BMP \mathbf{U}$ fits $\mathbf{X}$ up to residuals. The Boolean matrices $\mathbf{X}$, $\mathbf{Z}$, and $\mathbf{U}$ encode the relations $\upa$, $\ua$, and $\pa$, respectively.
    \label{figure:dimensions} }
\end{figure}

\subsection{Assumptions and problem definition}
\label{rmDef_newDef}
\begin{figure}[htb]
  \centering
    \begin{minipage}[c]{0.5\linewidth}    
    \centering
\includegraphics[width=0.99\textwidth] {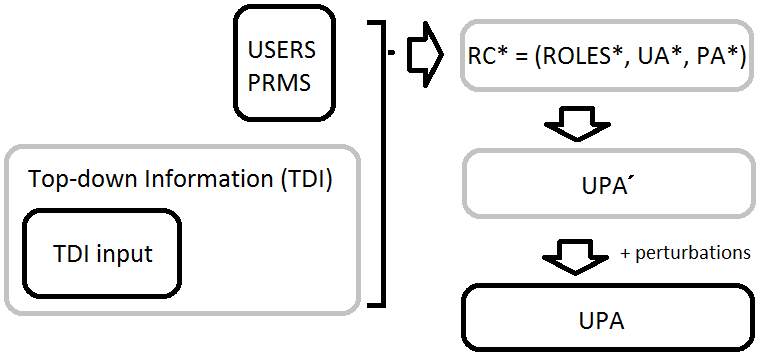}
    \end{minipage}
    \hfill
    \begin{minipage}[c]{0.49\linewidth}
      \caption{\label{rmDef_fig:DefScheme} Dependencies of the
        entities involved in role mining. This scheme illustrates how
        the user-permission assignment $\upa$ is assumed to be
        generated. An arrow indicates the direction of the generation
        process. Grey entities are unknown and black ones are given as
        input. For pure bottom-up role mining, no top-down information
        is given.}
    \end{minipage}
\vspace{-1em}
\end{figure}

The role mining problem is defined in terms of a set of users $\user$,
a set of permissions $\perm$, a user-permission assignment relation
$\upa$, {a set of roles $\roles$, a user-role assignment relation $\ua$, a role-permission assignment relation $\pa$,} and, if available, top-down information $\tdi$.  Our problem
definition is based on three assumptions about the relationships
between these entities and the generation process of $\upa$.  All
these entities and their relationships  are sketched
in Figure~\ref{rmDef_fig:DefScheme}.

\paragraph{Assumption 1: An underlying RBAC configuration exists}
  We assume that the given user-permission assignment $\upa$ was
  induced by an unknown RBAC configuration
  $RC^*\!\!=\!(\roles^*\!\!,\ua^*\!\!,\pa^*)$, where ``induced'' means
  that $\upa\approx\ua^*\BMP \pa^*$.  This assumption is at the heart
  of role mining.  To search for roles implicitly assumes that they
  are there to be found.  Said another way, searching for roles in
  direct assignments between users and permissions only makes sense if
  one assumes that parts of the data could, in principle, be organized
  in such a structured way.  Without this assumption, the role
  structure that one can expect to find in a given access-control
  matrix would be random and therefore meaningless from the
  perspective of the enterprise's business processes and security
  policies.

  \paragraph{Assumption 2: Top-down information $\tdi$ influences the
    RBAC configuration $\rc^*$}
  We assume that $RC^*$ reflects the enterprise's security policies
  and the business processes in the sense that $RC^*$ encodes these
  policies and enables users to carry out their business tasks.  Full
  knowledge of the security policies and business processes, as well
  as all business attributes and user tasks, should, in principle,
  determine the system's user-permission assignment.  We denote all
  such information as top-down information (TDI); this name reflects
that the task of configuring RBAC using TDI is usually referred to as
  top-down role mining. In practice, only parts of TDI may be
  available to the role mining algorithm. In
  Figure~\ref{rmDef_fig:DefScheme}, we account for this {structure} by
  distinguishing ``TDI'' from ``TDI input''. Whenever some parts of
  TDI are used in the role mining process we then speak of hybrid-role
  mining.

  \paragraph{Assumption 3: Exceptions exist}
  The observed user-permission assignment $\upa$ might contain an
  unknown number of exceptions. An exception is a user-permission
  assignment that is not generated by the role-structure $RC^*$ but
  rather from a set of unknown perturbation processes. To capture this
  assumption in Figure~\ref{rmDef_fig:DefScheme}, we distinguish the
  given access control matrix $\upa\approx\ua^*\BMP \pa^*$ from an
  unknown exception-free matrix $\upa'=\ua^*\BMP \pa^*$ that is fully
  determined by the role structure.  We emphasize that, even though
  there are many ways errors {can arise} in a system, exceptions
  are not necessarily errors.  Moreover, a role mining algorithm
  cannot be expected to discriminate between an error (that is, an
  unintended exception) and an intentionally made exception, if
  additional information is not provided.  A role mining algorithm can
  identify exceptions and report them to a domain expert (ideally
  ranked {by} their likelihood). Such a procedure already
  provides a substantial advantage over manually checking all user-permission
  pairs as it involves far fewer checks.  Due to the lack of
  additional information, we abstain from making further assumptions
  about the exceptions, for instance the fraction of exceptions
  $\delta$. Instead, determining such parameters will constitute an
  important part of the role mining problem.
\medskip

We now propose
the following definition of the role mining problem.
\begin{definition}\textbf{Role inference problem} 
\label{rmDef_def_generalRM} 
\\
Let a set of users $\user$, a set of permissions $\perm$, a
user-permission relation $\upa$, and, optionally, parts of the
top-down information $\tdi$ be given. Under Assumptions 1--3, infer
the unknown RBAC configuration
$RC^*\!\!=\!(\roles^*\!\!,\ua^*\!\!,\pa^*)$.
 \end{definition} 

 This definition, together with the assumptions on $\upa$'s generation
 process, provides a unified view of bottom-up and hybrid role mining.
 The two cases only differ in terms of the availability of top-down
 information $\tdi$.  In hybrid role mining, parts of the top-down
 information that influenced $RC^*$ is available.  When $\tdi$
 {is not provided}, the problem reduces to bottom-up role mining.
 Note that in such cases the goal still remains the same: the solution
 to Problem~\ref{rmDef_def_generalRM} solves the bottom-up problem as
 well as the hybrid role mining problem.  Thereby, the assumption that
 $RC^*$ is (partially) influenced by $\tdi$ is also reasonable for the
 pure bottom-up role mining problem. Whether $\tdi$ actually
 influences $RC^*$ does not depend on the availability of such data.

\paragraph{Relationship to {role mining by compression}} 
The role mining problems that aim to achieve the closest fit for a
given compression ratio or achieve the best compression for a given
deviation differ from the role inference problem.  Technically, our
definition has less input {than the alternatives}. For the
above-mentioned problems, either the number of roles $k$ or the
deviation $\delta$ is provided as input.  In contrast, our definition makes no
assumptions on these {quantities} and both $k$ and $\delta$ must be
learned from $\upa$ (as finding $RC^*$ involves finding $k$ and, at
the same time, determines $\delta=\upa-\ua^*\BMP \pa^*$).

Moreover, we see it is an advantage that the assumptions of the
problem are explicitly given.  By making the assumption of an
underlying role structure a condition for role mining, our problem
definition favors conservative algorithms in the following sense. If
 little  or no structure exists in $\upa$, the optimal
algorithm should refrain from artificially creating too many roles
from $\upa$. In contrast, if the number of roles or the closeness of
fit is predetermined, optimal algorithms will migrate exceptional (and
possibly 
{unwarranted}) permissions to RBAC.

Finally, it is unrealistic {in practice} that $\delta$ or $k$ will be
given as  an input. {Hence}, 
treating them as unknowns reflects real-world scenarios 
better than previous definitions for role mining that require
either $\delta$ or $k$ as inputs.

\subsection{Quality measures}
\label{rmDef_qualmeas}
\paragraph{Comparison with true roles}
The obvious quality measure that corresponds to the role inference
problem is the distance to the hidden RBAC configuration $RC^*$
underlying the given data $\upa$.  Several distance metrics for
comparing two RBAC configurations exist, for instance the Hamming
distance between the roles or the Jaccard similarity between the
roles. Usually, however, $RC^*$ is not known in practice.
A comparison is possible only for artificially created user-permission
assignments where we know $RC^*$ or when an existing RBAC
configuration is used to compute a user-permission assignment.  We
therefore focus in this paper on quality measures that are applicable
to all access control matrices, independent of knowledge of $RC^*$.

\paragraph{Generalization error} 
We propose {to use generalization error for evaluating} RBAC
configurations.  The generalization error is often used to assess
supervised learning methods for prediction
\cite{Hastie:Tibshirani:Friedman:2001}.  The generalization error of
an RBAC configuration $RC$ that has been learned from an input dataset
$\mathbf{X}^{(1)}$ indicates how well $RC$ fits to a second dataset
$\mathbf{X}^{(2)}$ that has been generated in the same way as
$\mathbf{X}^{(1)}$.

Computing the generalization error for an unsupervised learning
problem like role mining is conceptually challenging.  
{In general,}
it is unclear how to transfer the {inferred} roles to a hold-out test
dataset when no labels are given that indicate a relationship between
roles and users.  We employ a method that can be used for a wide
variety of unsupervised learning problems: the transfer costs proposed
in \cite{mtc_ECML}.
The transfer costs of a role mining algorithm are computed {as follows}.  First the input dataset is randomly split along the
users into a training set $\mathbf{X}^{(1)}$ and a validation set
$\mathbf{X}^{(2)}$. Then the role mining algorithm learns the RBAC
configuration $RC=(\hat{\mathbf{Z}}$, $\hat{\mathbf{U}})$ 
based {only} on the training set {and without} any 
knowledge of the second hold-out dataset.  Having learned $RC$, the
solution is transfered to the hold-out dataset by using a nearest
neighbor mapping between the users of both datasets.  Each user in
$\mathbf{X}^{(2)}$ is assigned to the set of roles of its nearest
neighbor user in $\mathbf{X}^{(1)}$. Technically, this means we keep $\hat{\mathbf{U}}$
fixed and create a new assignment matrix $\mathbf{Z}'$, where row $i$
is copied from the row in $\hat{\mathbf{Z}}$ that corresponds to the
nearest neighbor user of user $i$.  Then generalization error is the
Hamming distance between $\mathbf{Z}' \BMP \hat{\mathbf{U}}$ and
$\mathbf{X}^{(2)}$ divided by the number of entries of
$\mathbf{X}^{(2)}$. This {ratio denotes} the fraction of {erroneously}  generalized
assignments.

The rationale behind our measure is intuitive: Since the input dataset
is assumed to be generated by an unknown RBAC configuration $RC^*$,
the subsets $\mathbf{X}^{(1)}$ and $\mathbf{X}^{(2)}$ have also been
generated by $RC^*$. The structure of the input matrix is the same in
different subsets of the dataset, but the random exceptional
assignments are unique to each user.  A role mining algorithm that can
infer $RC^*$ from one subset, will have a low generalization error
because the generating RBAC configuration should generalize best to
the data that it has generated.  As a consequence, a role mining
algorithm that overfits to noise patterns in one dataset will fail to
predict the structure of the second dataset.  Thereby, even a perfect role mining algorithm will have a positive generalization error if the
data is noisy. It is the relationship to the generalization error of other algorithms that counts. All methods fail to predict exceptional
assignments but the better methods will succeed in identifying the structure of the hold-out data while the inferior methods {will compromise the underlying structure to adapt to exceptions}.

{ An advantage of computing the generalization error as
  described above is that this computation is agnostic to the role   mining algorithm used. In particular it works for both probabilistic   and combinatorial methods.} {For probabilistic methods one could achieve improved results by using a posterior inference step to assign test users to the roles discovered.} {However, this method is tailored to the
  particular methods {employed} and would not work for methods without a
  probabilistic model.  }

\section{From a deterministic rule to a class of probabilistic models} 
 \label{sec_probmodels}

 In this section we propose a class of probabilistic models for role
 mining. We derive our core model from the deterministic assignment rule
 of RBAC and extend this core model to more sophisticated models.  We
 will present two such extensions: (1) the disjoint-decomposition model
 (DDM) with a \emph{two-level role hierarchy} where each user has only
 one role and (2) multi-assignment clustering (MAC), a \emph{flat RBAC}
 model featuring a role relationship where users can assume multiple
 roles.  We will {also} show why these two instances of the model class are
 particularly relevant for role mining.

\subsection{Core model} 
\label{sec_models}
In the following, we derive the core part of our probabilistic model.
We start with the deterministic rule that assigns users to permissions
based on a given role configuration. We then convert this rule into a
probabilistic version, where one reasons about the probability of
observing a particular user-permission assignment matrix given the
probabilities of users having particular roles and roles entailing
permissions.

We denote the user-permission assignment matrix by $\mathbf{X}$, with
$\mathbf{X} \in \{0,1\}^{N\times D}$. As short-hand, we write
$\mathbf{x}_{i\sep \allInd}$ for the $i^{\textrm{th}}$ row of the
matrix and $\mathbf{x}_{\allInd\sep d}$ for the
$d^{\textrm{th}}$ column.  We define the generative process of {an
  assignment} $x_{id}\in\left\{0,1\right\}$ by
\begin{eqnarray}
u_{kd} & \sim & p(u_{kd} \vert \beta_{kd}) \\
x_{id} & \sim & p(x_{id}\vert \mathbf{u}_{\allInd \sep d}, \mathbf{z}_{i\sep \allInd}) \,,
\end{eqnarray}
where $a \sim p(a)$ means that $a$ is a random variable drawn from the
probability distribution $p(a)$. 
The latent variable $u_{kd}\in\left\{0,1\right\}$ determines {permission} ${d}\in\left\{1,\myldots,D\right\}$ of source $k\in\left\{1,\myldots,K\right\}$. The parameter $z_{ik}\in\left\{0,1\right\}$ encodes whether {user} $i$ is assigned to {role} $k$. 
{As $u_{kd}$ is binary, $p(u_{kd} \vert \beta_{kd})$ is a Bernoulli
  distribution}, with $p(u_{kd}=0):=\beta_{kd}$ {and}
\begin{equation}
p(u_{kd}\vert \beta_{kd}):=\beta_{kd}^{1-u_{kd}} (1-\beta_{kd})^{u_{kd}} \,.
\label{eq_pOfU}
\end{equation}
Throughout this section, we will condition all probabilities on
$\mathbf{Z}$. Therefore, we can ignore $p(z_{ik})$ for the moment and
treat it as a model parameter here. In Section~\ref{DDMmodel} we will
treat $z_{ik}$ as a random variable and describe a particular prior
distribution for it.

\begin{figure}[htb]
  \centering
    \begin{minipage}[c]{0.6\linewidth}    
    \centering
\includegraphics[width=0.37\textwidth]{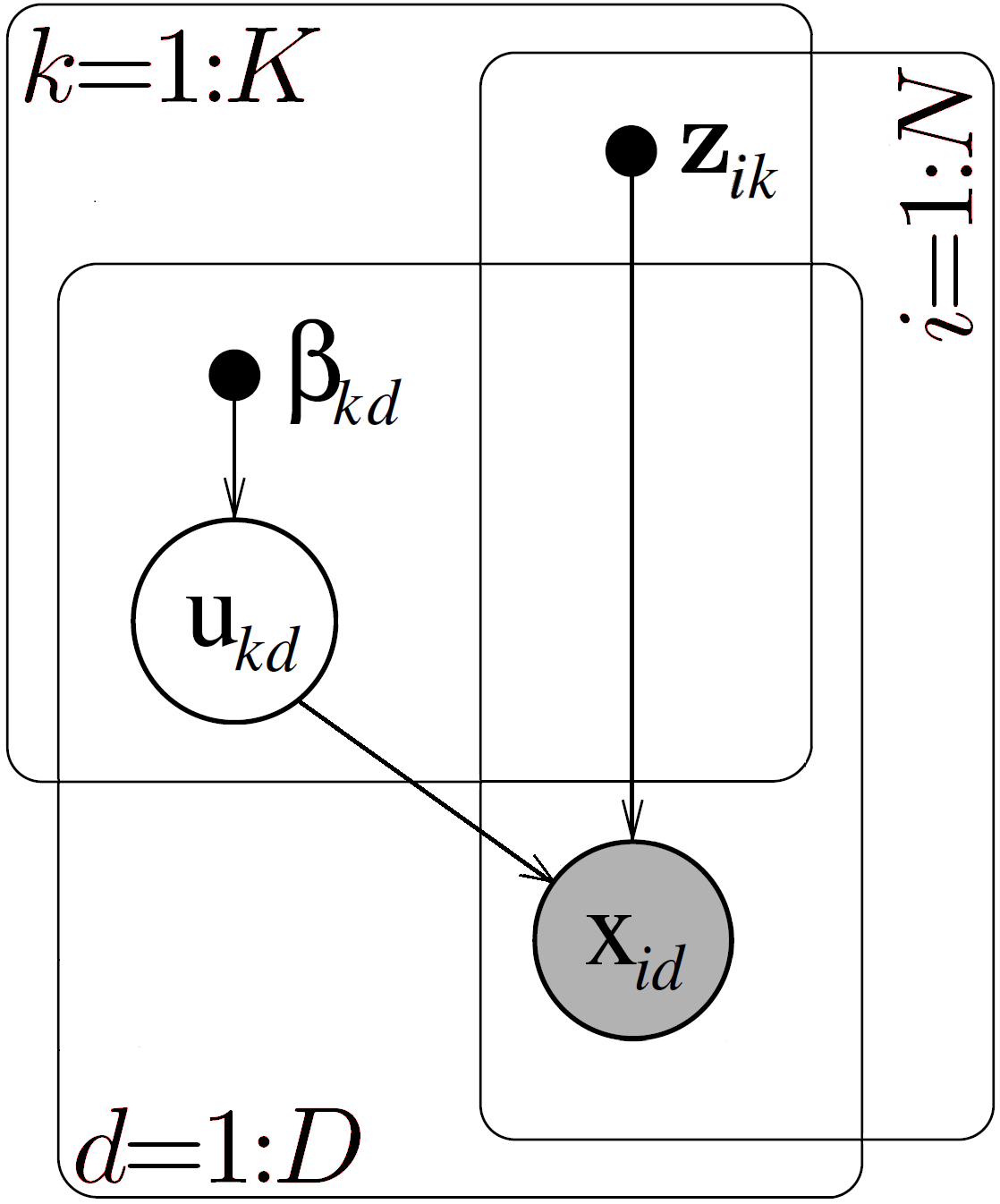}
    \end{minipage}
    \hfill
    \begin{minipage}[c]{0.39\linewidth}
     \caption{\label{fig:genModel_basic}  Graphical model corresponding
       to the generation rule of user-permission assignments given an
       RBAC configuration. See the explanation of the semantics in the text.  
}
    \end{minipage}
\end{figure}

The generative model that we described so far is illustrated in
Figure~\ref{fig:genModel_basic}. {All entities in this figure
 have a {well-defined semantic} meaning.} The circles are random variables. 
A filled circle denotes that
the variable is observable (like the user-permission
assignment $x_{id}$) and an empty circle represents a hidden variable.
Small solid dots are unknown model parameters. Arrows indicate
statistical dependencies between entities. Whenever an entity is in a
rectangle (say in the $N$-rectangle and in the $D$-rectangle) then multiple
different realizations of this entity exist (here $N\cdot D$
realizations of $x_{id}$, each with a different index $i$ and $d$).

The probability $p(x_{id}\vert \mathbf{u}_{\allInd \sep d}, \mathbf{z}_{i\sep \allInd})$ is deterministic in the following sense. Given all {role-permission assignments} $\mathbf{u}_{kd}$ and the {role assignments} $\mathbf{z}_{i\sep \allInd}$, the bit $x_{id}$ is determined by the  disjunction rule defined by the Boolean matrix product {
\begin{equation}
\label{eq_xEqzu}
 \mathbf{X}= \mathbf{Z} \BMP \mathbf{U} \ \ \ \text{such that} \ \ \ \ \ \ x_{i \sep d}=f(\mathbf{u}_{\allInd \sep d}, \mathbf{z}_{i\sep
   \allInd}):=\bigvee_{k = 1}^K \left(z_{i \sep k}\wedge u_{k \sep d}\right) \ .
 \end{equation} }
 To derive the likelihood of $x_{id}$, we express this deterministic
 formula $x_{id}=f(\mathbf{u}_{\allInd \sep d}, \mathbf{z}_{i\sep
   \allInd})$ in terms of a probability distribution. To this end, 
    the entire probability mass must be concentrated at the
 deterministic outcome, i.e. the distribution must be of the form
\begin{equation}
p(x_{id}\vert \mathbf{u}_{\allInd \sep d}, \mathbf{z}_{i\sep \allInd}) =
\left\{\begin{array}{cl} 1, & \mbox{if }x_{id}=f(\mathbf{u}_{\allInd
      \sep d}, \mathbf{z}_{i\sep \allInd})\\ 0, &
    \mbox{otherwise} \end{array}\right. \,.
\end{equation}
A probability distribution that fulfills this requirement is
\begin{equation}
\label{eq_bitGenerationDeterm}
p(x_{id}\vert \mathbf{u}_{\allInd \sep d}, \mathbf{z}_{i\sep \allInd}) = \Bigl(\prod_k\left(1-u_{kd}\right)^{z_{ik}}\Bigr)^{(1-x_{id})} \Bigl(1-\prod_k\left(1-u_{kd}\right)^{z_{ik}}\Bigr)^{x_{id}}.
\end{equation}

This can be seen by going through all eight
combinations (for a single $k$) of the binary values of $z_{ik}$,
$x_{id}$, and $u_{kd}$.  The distribution reflects that there are only two
possible outcomes of this random experiment. In the following, we exploit this
property and work just with probabilities for $x_{id}=0$. The probability
for $x_{id}=1$ is always  
the remaining
probability mass.

The model in its current deterministic state is not directly useful for
role mining given the hidden variables $u_{kd}$. We therefore 
eliminate the $u_{kd}$ to obtain a likelihood that only depends on the
model parameters and the observations. This can be achieved by
marginalizing out $u_{kd}$, i.e., summing over all possible $K
\times D$ matrices $\mathbf u$.
  As derived in
Appendix~\ref{app_derivLikelih}, this yields the
likelihood
$
p(x_{id}=0 \vert  \boldsymbol \beta_{\allInd \sep d}, \mathbf{z}_{i\sep \allInd}) = 
 \prod_{k}  \beta_{kd}^{z_{ik}} 
$. 
{This term reflects that if a user is not assigned to a role, {then} the role {does not have any} influence on the user's permissions.}  There{fore,} the chance of
a user not being assigned some permission $d'$ decreases with the number of roles of the user, since the chances
$\beta_{kd'}$ of $d'$ not being assigned to the roles are multiplied.

As $x_{id}$ can only take two possible values, we have $p(x_{id}=1
\vert \boldsymbol \beta_{\allInd \sep d}, \mathbf{z}_{i \sep \allInd})
= 1-\prod_{k} \beta_{kd}^{z_{ik}} $ such that the full likelihood of
the bit $x_{id}$ is
\begin{equation}
  p(x_{id} \vert \boldsymbol \beta_{\allInd \sep d}, \mathbf{z}_{i
    \sep \allInd}) = \left(\prod_{k}
    \beta_{kd}^{z_{ik}}\right)^{1-x_{id}} \left(1-\prod_{k}
    \beta_{kd}^{z_{ik}}  \right)^{x_{id}} \ . 
\label{eq_mainDataLikelihood}
\end{equation}
According to this likelihood, the different entries in $\mathbf{X}$
are conditionally independent given the parameters $\mathbf{Z}$
and $\boldsymbol{\beta}$. Therefore, the complete data likelihood
factorizes over {users and permissions}:
$\
  p(\mathbf{X} \vert \boldsymbol{\beta}, \mathbf{Z}) = \prod_{i=1}^N \prod_{d=1}^D
  p(x_{id} \vert \boldsymbol \beta_{\allInd \sep d}, \mathbf{z}_{i \sep \allInd})
$.

\subsection{Role hierarchies} \label{prob-model} 
 In this section we extend the core model by introducing
role hierarchies. The core model provides a hierarchy of depth 1 as
there is only one level of roles. We introduce an additional level,
resulting in a hierarchy of depth 2. The meaning of the hierarchical
relationship is as follows. Roles in the second layer can be sub-roles
of roles in the first layer. The set of permissions for a super-role
includes all permissions of its sub-roles.

Our derivation is generic in that it can be used to add extra layers to a
hierarchical model. By repeated application, one can derive
probabilistic models with hierarchies of any depth. As we will see, the one-level
hierarchy (flat RBAC) and the two-level hierarchy are particularly 
interesting. In Section~\ref{sec_instantiation} we propose a model
variant for flat RBAC and a model variant with a two-level hierarchy.

Like the core model, the hierarchical model derived here is not restricted to role mining. However, we will motivate {its} usefulness of
hierarchies by practical considerations in access-control.  We assume
that there exists a decomposition of the set of users into partially
overlapping groups: Users are assigned to one or more groups by a
Boolean assignment matrix $\mathbf{Z}$. Each row $i$ represents a user
and the columns $k$ represent user-groups. In practice, such a
decomposition may be performed by an enterprise's Human Resources Department, for example, by assigning users to {enterprise} divisions according to defined similarities of the employees. If such
data is lacking, then the decomposition may just be given by the
differences in the assigned permissions for each user. For simplicity,
the matrix $\mathbf{Z}$ has the same notation as in the last section.

We {now} introduce a second layer. We assume that there is a decomposition
of the permissions such that every permission belongs to one or more
permission-groups. These memberships are expressed by the Boolean
assignment matrix $\mathbf{Y}$. Here the $l$th row of $\mathbf{Y}$
represents the permission-group $l$ and the $d$th column is the
permission $d$. The assignment of permissions to permission-groups can
be motivated by the technical similarities of the resources that the
permissions grant access to.  For example, in an object-oriented
setting, permissions might be grouped that execute methods in the same
class. Alternatively, permissions could be categorized based on the
risk that is associated with granting someone a particular
permission. Of course, permissions can also be grouped according to
the users who own them.

We denote user-groups by \textbf{business roles} whereas we call
permission-groups \textbf{technical roles}.  Business roles are
assigned to technical roles. We represent these assignments by a
matrix $\mathbf{V}$.  To keep track of all introduced variables, we
list the types of the above-mentioned Boolean assignment matrices:

\vspace{1ex}
\begin{tabular} {cll}
$\bullet$ & Users $i$ to permissions $d$: & $ \!x_{id}\!\in\!\left\{0,1\right\}$, where
  $i\!\in\!\left\{1,\myldots,\!N\right\},
  d\!\in\!\left\{1,\myldots,\!D\right\}$. \\
$\bullet$ &  Users $i$ to business roles $k$: & $z_{ik}\in\left\{0,1\right\}$, where
  $k\in\left\{1,\myldots,K\right\}$. \\
$\bullet$ & Business roles $k$ to technical roles $l$: &
$v_{kl}\in\left\{0,1\right\}$,
where $l\in\left\{1,\myldots,L\right\}$.\\
$\bullet$ & Technical roles $l$ to permissions $d$: & $y_{ld}\in\left\{0,1\right\}$.
\vspace{1ex}
\end{tabular} 
Throughout this section, the indices $i$, $d$, $k$, and $l$ have
the above scope and are used to index the above items. 

Starting with this additional layer of roles, one can recover a flat
hierarchy by collapsing the role-role assignment matrix and the
role-permission assignments using the disjunction rule 
$
u_{kd} = \bigvee_{l} v_{kl}\wedge y_{ld} 
$.
Thereby,  $\mathbf{U}$ can be understood as the role-permission
assignment matrix from the last section. 
With this {structure}, the final $N\times D$
user-permission assignment matrix $\mathbf{X}$ is determined by two
Boolean matrix products 
\begin{align}  
&\mathbf{X}= \mathbf{Z}\BMP \mathbf{U}
  = \mathbf{Z}\BMP \mathbf{V}\BMP \mathbf{Y} 
  \label{boolmatprod}
  \ \ \
  \text{ with }~
  x_{id}=\bigvee_{k}\left[z_{ik}\wedge\left(\bigvee_{l}v_{kl}\wedge
  y_{ld}\right) \right] \, . 
\end{align}

Equation~\ref{boolmatprod} expresses {when} a user $i$ is assigned to  
permission $d$. There {exists} one Boolean matrix product per role layer.
{Note that for a given RBAC configuration, we can also
  partially collapse hierarchies. In particular, this makes
  sense when a business role is directly linked to permissions.} 
We are again interested in
the probability of such an assignment. Starting from this logical
expression, we derive below how likely it is to observe an assignment of a
user $i$ to a permission $d$.

The deterministic assignment rule for two layers of roles is
graphically illustrated in Figure~\ref{figModels}(a):  a user is
assigned to a permission if there is at least one path in the graph 
connecting them.   As  this figure indicates, a user can be assigned to a
permission in multiple ways, that is, there may be multiple paths.
It is therefore easier to express how a user may \emph{not} be
assigned to a permission (we denote this by $\neg x:=\overline{x}$)
rather than computing the union over all possible assignment paths. {Also, we will abbreviate  parameters by $z^{+}\!:=\!p(z\!=\!1)$ and, for independent variables, $\mathbf{z}_{i  \sep \allInd}^{+}\!:=\!\{ z^{+}_{i  \sep 1},\myldots, z^{+}_{i  \sep K}  \}$.}
\begin{eqnarray}
\llap{$p$}\left(\overline{x_{id}}\  {\vert\mathbf{z}_{i  \sep \allInd}^{+} ,\mathbf{y}_{ \allInd \sep d}^{+}, V^{+}} \right) 
&=&
p\left(\overline{\bigvee_{ k=1}^K \left[z_{ik} \wedge
    \left(\bigvee_{l=1}^L v_{kl}\wedge y_{ld}\right) \right]} \
{\vert \mathbf{z}_{i  \sep \allInd}^{+},\mathbf{y}_{ \allInd \sep d}^{+}, V^{+}}\right)      
    \label{eq_2layerRule_short}
\end{eqnarray} 
As shown in Appendix~\ref{app_hierarchy}, this probability is
\begin{equation}  
  \llap{$p$}\left(\overline{x_{id}}\  {\vert\mathbf{z}_{i  \sep \allInd}^{+} ,\mathbf{y}_{ \allInd \sep d}^{+}, V^{+}} \right) 
 =
{ \prod_{k}
\left(
1-z_{i  \sep k}^{+}
 +z_{i  \sep k}^{+}\  
\prod_{l}
  \left(
 1- y_{ l \sep d}^{+})
+y_{ l \sep d}^{+} (1- v_{ k \sep l}^{+})  
\right) 
\right) }\, .
\end{equation} 
We condition this expression on the binary entries of $\mathbf{Y}$ and $\mathbf{Z}$.
\begin{eqnarray}
  p(\overline{x_{id}}\left.\right| \mathbf z_{i\sep \allInd}, \mathbf y_{\allInd \sep d} {, v_{ k \sep l}^{+}} ) &\!=\!&
  \prod_{k}1^{1-z_{ik}}\cdot\left(\prod_{l}
  {(1- v_{ k \sep l}^{+})}^{y_{ld}}\cdot1^{1-y_{ld}}\right)^{z_{ik}} \!\!
= \!
  \prod_{k,l}{(1- v_{ k \sep l}^{+})}^{z_{ik}y_{ld}} 
\end{eqnarray}

As this expression is independent of other matrix entries in
$\mathbf{X}$, we can express the complete likelihood of the
user-permission assignment matrix given the business roles and
technical roles as a product over users and permissions. 
\begin{align} \label{eq_fullmodel}
 p\left(\mathbf{X}\left.\right|\mathbf{Z},\mathbf{Y}\right)  
 &=
  \prod_{i,d}
  \left[1-p(\overline{x_{id}}\left.\right| \mathbf z_{i\sep \allInd}, \mathbf y_{\allInd \sep d} {, v_{ k \sep l}^{+}} ) \right]^{x_{id}}
  \left[p(\overline{x_{id}}\left.\right| \mathbf z_{i\sep \allInd}, \mathbf y_{\allInd \sep d} {, v_{ k \sep l}^{+}} ) \right]^{1-x_{id}}
  \nonumber \\  
&= 
  \prod_{i,d} \left[1-\prod_{k,l}{(1- v_{ k \sep l}^{+})}^{z_{ik}y_{ld}}
    \right]^{x_{id}} \left[
    \prod_{k,l}{(1- v_{ k \sep l}^{+})}^{z_{ik}y_{ld}}\right]^{1-x_{id}} 
\end{align} 

If we treat { $v_{ k \sep l}$ as a random variable with probability ${p(v_{ k \sep l}=0)=(1- v_{ k \sep l}^{+})}$}, then this likelihood
resembles the one with only one layer of roles. The only differences
are the additional binary variables $y_{ld}$ in the exponent that,
like $z_{ik}$, can switch off individual terms of the
product. Thereby, we chose to condition on $\mathbf{Y}$ and
$\mathbf{Z}$ and leave $v_{kl}$ random.  We could as well have
conditioned on $\mathbf{V}$ and $\mathbf{Z}$ and inferred $y_{ld}$.
This alternative points to a generic inference
strategy for role hierarchies of arbitrary size. One treats the
assignment variables in one layer as a random variable and conditions on
the current state of the others. We will demonstrate such an alternating
inference scheme on a two level hierarchy in Section~\ref{DDMmodel}.

\subsection{Overparametrization and instantiation by introducing constraints}
\label{sec_instantiation}
The above model of user-permission assignments
defines a very general framework.  In principle, one can iteratively
introduce additional layers in the role hierarchy without substantially
changing the outer form of the likelihood.
In the derivations, we have avoided any 
prior assumptions on the probabilities of the entries of $\mathbf{V}$,
$\mathbf{Y}$, and $\mathbf{Z}$. We have only exploited the fact that
these variables are Booleans and, therefore, only take the values
0 or 1. We have also avoided any assumptions about the processes that
lead to a particular decomposition of the set of users and the set of
permissions. Moreover, we have not specified any constraints
on the user decomposition, the permission decomposition, or the
assignments from user-groups to permission-groups.

It turns out that the model with a two-level hierarchy already has
more degrees of freedom than is required to represent the access control
information present in many domains that arise in practice.  In
particular, when only the data $\mathbf{X}$ is given, there is no
information available on how to decompose the second role level. This 
{lack of identifiability} becomes obvious
when we think about a one-level decomposition with role-permission
assignments $\mathbf{U}$ (as in Eq.~(\ref{eq_xEqzu})) and try to
convert it into a two-level decomposition. The $\mathbf{Z},\mathbf{U}$
decomposition has already sufficiently many degrees of freedom to fit
any binary matrix. Further decomposing $\mathbf{U}$ into an extra
layer of roles $\mathbf{V}$ and assignments from these roles to
permissions $\mathbf{Y}$ is arbitrary when there does not exist additional
information or constraints.  Therefore, the flat RBAC configuration
with only one role layer is the most relevant one.  The two-level
hierarchy without constraints is over-parameterized.  Such a hierarchy
can be seen as a template for an entire class of models. By
introducing constraints, we can instantiate this template to
specialized models that fit the requirements of particular RBAC
environments and have a similar model complexity as flat RBAC without
constraints. These instances of the model class are given by
augmenting unconstraint two-level RBAC with assumptions on the
probability distributions of the binary variables and giving
constraints on the variables themselves. In the following, we will
present two relevant model instances and explain their relationship.
Later, we will extend the models with generation processes for
exceptional assignments.

\subsubsection*{Flat RBAC}
\begin{figure}[tb]
\centering
\includegraphics[width=1\textwidth]{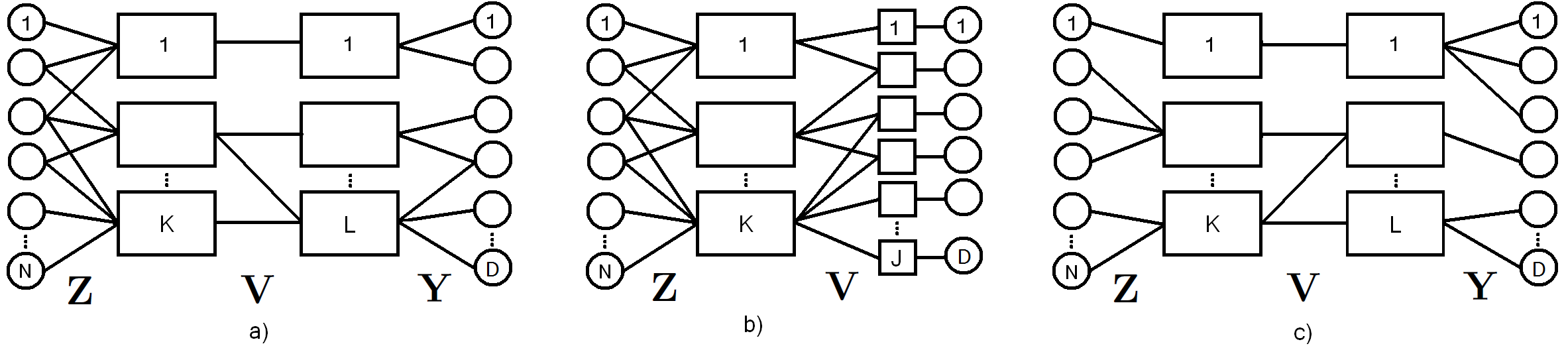}	
\caption{Illustration of the structure of three model instances. A
  user has a permission if there is at least one path connecting
  them. a) Full model.  b) Model with trivial decomposition of the
  permissions (flat RBAC). c) Disjoint Decomposition Model with only
  one business role per user and one technical role per permission.
  \label{figModels}}
\end{figure}

In this model, each permission is restricted to be a member of only
one permission-group and each permission-group can contain only a
single permission. Formally:
$
\textstyle \left( \forall
    l:\sum_{j}y_{ld}=1\right) \wedge \left( \forall j:
    \sum_{l}y_{ld}=1 \right) 
    $.
The conditioned likelihood then becomes 
\begin{equation} 
p_{\flatRBAC}\left(\mathbf{X}\left.\right|\mathbf{Z}\right) =
\prod_{i,d}\left[1-\prod_{k}p(\overline{v_{kd}})^{z_{ik}}
  \right]^{x_{id}} \left[\prod_{k}p(\overline{v_{kd}})^{z_{ik}} 
\right]^{1-x_{id}}\rlap{.}
\label{eq_flatRBAC}
\end{equation}
This ``collapsed'' model is equivalent to flat RBAC without
constraints. {This can be seen} by renaming $v$ by $u$ and $J$ by
$D$. As each technical role serves as a proxy for exactly one
permission, we have $D=J$ anyway. A graphical representation of the
structure of this model instance is given in
Figure~\ref{figModels}(b).  Equivalently, we could introduce
one-to-one constraints on the user role assignment and collapse
$\mathbf{Z}$ instead of $\mathbf{Y}$, leading to a model with the same
structure.

\subsubsection*{Disjoint decomposition model {(DDM)}}\label{DDMmodel}
{This} model has even stronger constraints.  Namely, $k_{max}=1$ and
the number of assigned permission-groups per permission is limited to
$l_{max}=1$.  This formalizes that each user belongs to exactly one
user-group and each permission belongs to exactly one
permission-group.  Hence, both users and permissions are respectively
partitioned into disjoint business roles and technical roles. A
disjoint decomposition substantially reduces the complexity of a
two-level hierarchy while still retaining a high degree of flexibility
since users of a given user-group may still be assigned to multiple
permission-groups.  We illustrate this model in
Figure~\ref{figModels}(c).

\subsection{Prior assumptions on probabilities}  
\label{eq_PriorAssumption}  
A central question in role mining is how many roles are required to 
explain a given dataset. 
In this paper, we take two different approaches to determining the
number of roles $k$.  For the flat RBAC model, we treat $k$ as a fixed
model parameter. One must therefore repeatedly run the
algorithm optimizing this model for different $k$ and select the
result according to an external measure. In our experiments, we will tune {the number of roles} $k$ by cross-validation using the generalization error as the external quality measure.  For {DDM} with a
two-level hierarchy, we explicitly include prior assumptions on the number of roles into the model using a non-parametric Bayesian
approach. This way, the role mining algorithm can internally select the number of roles.  

Instead of directly providing the number of roles for DDM, we only
assume that, given an RBAC configuration with roles and user-role
assignments, the a priori probability of a new user being assigned to one of the roles depends linearly on the number of users having this
role (plus a nonzero probability of creating a new role). This assumption reflects that it is favorable to assign a new user to
existing roles instead of creating new roles when users enter the enterprise.  This rich-get-richer effect indirectly influences the
number of roles as, under this assumption, an RBAC configuration
with few large roles (large in the number of users) is more likely
than a configuration with many small roles.

Our assumption is modeled by a Dirichlet process prior
\cite{antoniak:dp,ferguson:dp}. Let $N$ be the total number of users and
let $N_k$ be the number of users that have role $k$ (the cardinality of the role).  The Dirichlet process is parametrized by the nonnegative
\textbf{concentration parameter} $\alpha$.
\begin{eqnarray}
\label{eq_Dirprior}
p\left(z_{i'k}=1\vert\mathbf{Z}_{i\neq i'}, \alpha\right) =
\begin{cases} \frac{N_k}{N-1+\alpha} & N_k>0\\
\frac{\alpha}{N-1+\alpha}& N_k=0
\end{cases} 
\end{eqnarray}
Here, we used the short-hand notation $\mathbf{Z}_{i\neq i'}$ for the
role assignments of all users except (the hypothetically new) user
$i'$. The event where user $i'$ is assigned a role with $N=0$
corresponds to creating a new role with exactly the permissions of $i'$.

The business-role to permission-role assignments are distributed according to a Bernoulli distribution  
$
p\left( v_{kl} \right. \left.\vert\;
\mathbf{Z},\mathbf{Y},\boldsymbol{\beta}\right) =
\beta_{kl}^{1-v_{kl}} (1-\beta_{kl})^{v_{kl}} \,.
$ 
Again, the model parameter $\beta_{kl}$ accounts for the probability that the assignment is not active. We introduce a symmetric Beta prior for $\beta_{kl}$:
\begin{align}
P_b\left(\beta_{kl}; \gamma,\gamma \right) 
&:= 
\frac{\Gamma(2\gamma)}{2 \Gamma(\gamma)} \left(\beta_{kl} (1-\beta_{kl}) \right)^{\gamma-1}
= B(\gamma,\gamma)^{-1} \left(\beta_{kl} (1-\beta_{kl}) \right)^{\gamma-1} \,, 
\end{align}
where $B(.,.)$ and $ \Gamma(.)$ are the beta function and the gamma
function, respectively. We 
derive the update equations for a Gibbs sampler on this model  in
Appendix~\ref{app_eviDDM}. 

\begin{figure}[htb]
  \centering
    \begin{minipage}[c]{0.6\linewidth}    
    \centering
\includegraphics[width=0.55\textwidth]{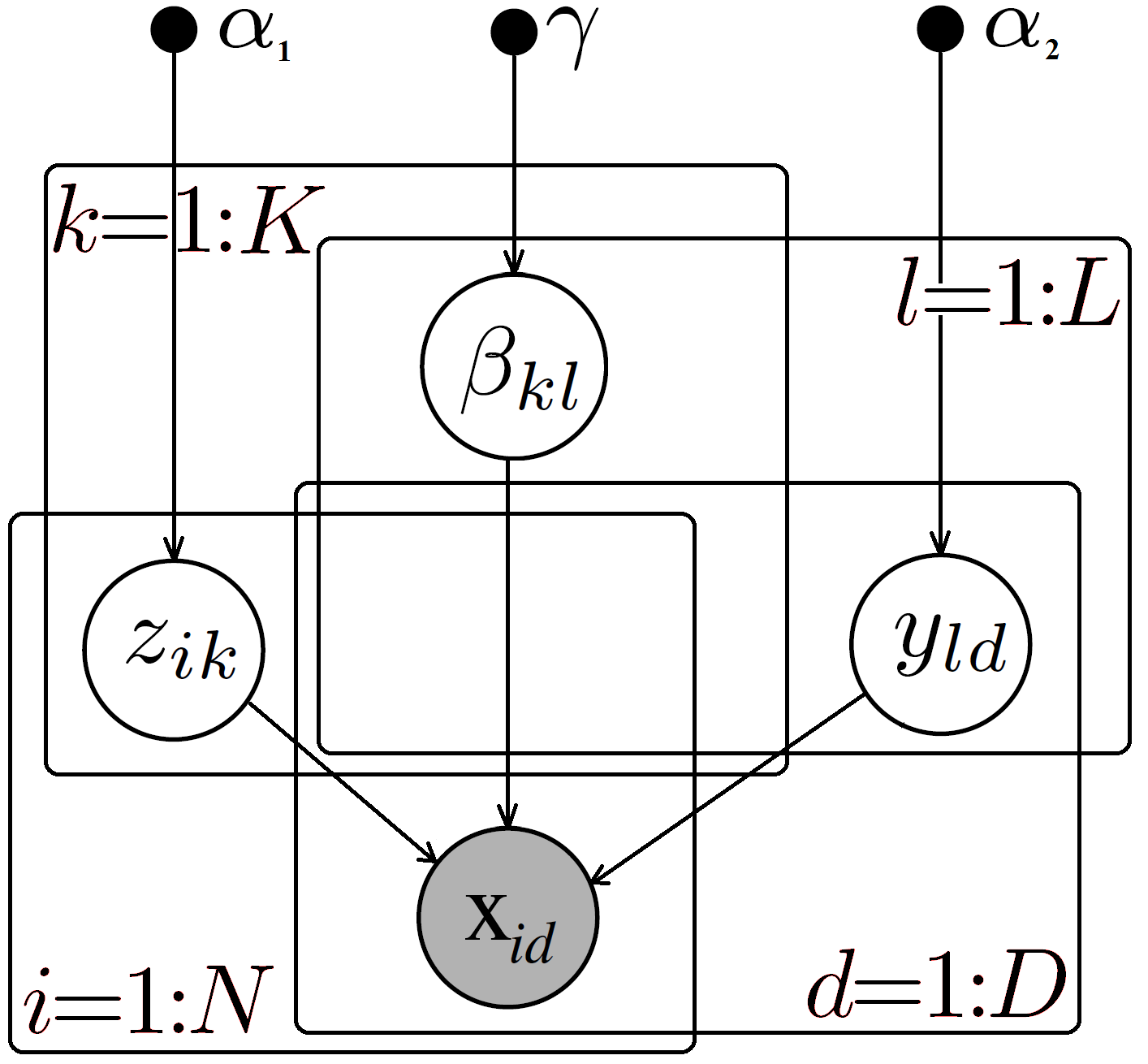}	
    \end{minipage}
    \hfill
    \begin{minipage}[c]{0.39\linewidth}
  \caption{\label{fig_DDMgraphmod}Graphical model for DMM with prior
  distributions.  
{Filled
  circles are {observed} random variables and empty circles are
  hidden. Solid points denote fixed parameters of the model. Arrows
  indicate dependencies. Entities on a $N$-rectangle exist in $N$ different realizations.  }
}
    \end{minipage}
\end{figure}

With the Dirichlet process prior on the user-role and the
role-permission  assignments and the Beta-Bernoulli likelihood, DDM
is equivalent to
the ``infinite relational model'' (IRM) proposed in \cite{Kemp}. We will
use a similar Gibbs-sampling algorithm to infer the model parameters of
DDM as used in \cite{Kemp}. 
While \cite{Kemp} use hill climbing to infer the model parameters, we
repeatedly sample the parameters from the respective probabilities and
keep track of the most likely parameters. 
We graphically illustrate the model with all introduced prior
assumptions in Figure~\ref{fig_DDMgraphmod}. There, we use different
hyperparameters $\alpha_1$ and $\alpha_2$ for the assignments
$\mathbf{Z}$ and $\mathbf{Y}$. However, in all calculations  we use the same hyperparameters $\alpha=\alpha_1=\alpha_2$.

Nonparametric Bayesian priors could, in principle, also be added to the
flat RBAC model proposed above. However, as this model allows users to
have multiple roles, the Dirichlet process prior is not applicable to
$\mathbf{Z}$. Instead, the 
``Indian Buffet Process'' \cite{IBP} could be
used as a prior. In our experiments, we compare with an algorithm that
combines such a prior with a noisy-OR likelihood \cite{IBPbinary}.

\subsection{Noise}
The goal of role mining is to infer the role structure underlying a
given access control matrix. In this context, structure means
permissions that are frequently assigned together and users that share
the same set of permissions (not necessarily all their
permissions). User-permission assignments that do not replicate over the
users do not account for the structure. We call such exceptional
assignments noise. A noisy bit can be an unusual permission or a missing assignment of a permission.

In this section we add an explicit noise process to the flat RBAC
model.  Let $x^S_{i\sep d}$ be a ``structural'' bit and let $x^N_{i\sep
  d}$ be a ``noise'' bit. While a structural bit is generated by the
generation process of the user-permission assignments in the flat RBAC
scenario given by Eq.~(\ref{eq_flatRBAC}), a noise bit is generated by a random
coin flip with probability $r$:
\begin{equation}
p_N \left(x^N_{i\sep d}\left.\right|r\right)
    = r^{x^N_{i\sep d}}\left(1-r\right)^{1-x^N_{i\sep d}}
    \label{eq:pNoise_xid_mix} \ .
\end{equation}

Let $\xi^S_{i\sep d}$ be a binary variable indicating if the observed
bit $x_{i\sep d}$ is a noise bit or a structure bit. Then the observed
bit is
\begin{equation}
  x_{i\sep d} = (1-\xi_{i\sep d}) x^S_{i\sep d} + \xi_{i\sep d} x^N_{i\sep d} \  .
\end{equation}

\begin{figure}[htb]
  \centering
    \begin{minipage}[c]{0.7\linewidth}    
    \centering
 \includegraphics[width=0.65\textwidth]{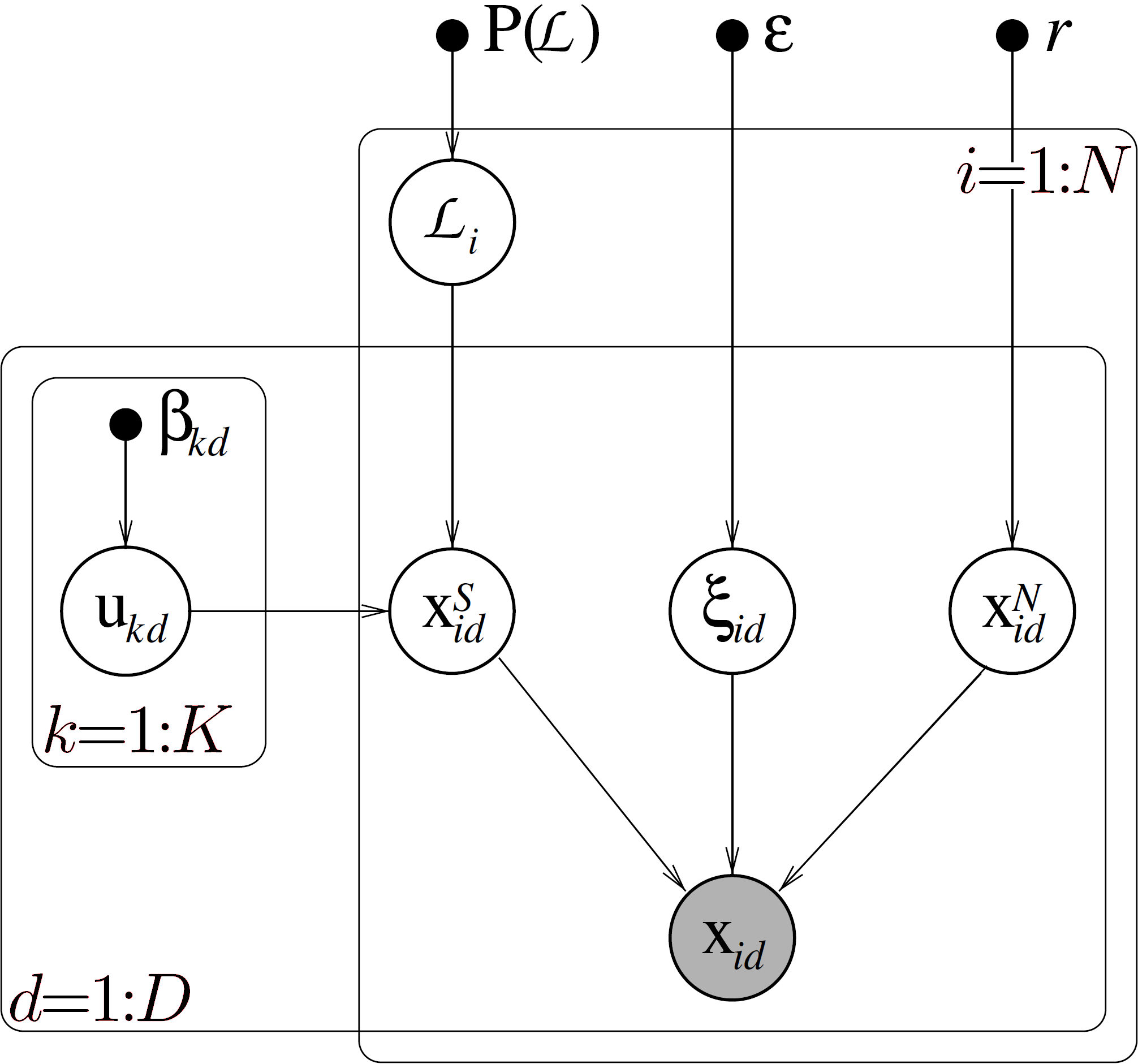}
    \end{minipage}
    \hfill
    \begin{minipage}[c]{0.29\linewidth}
  \caption{\label{fig:genModel} The generative model of Boolean MAC
   with mixture noise. $\lset_i$ is the assignment set of {user} $i$,
   indicating which {roles} from $\mathbf{U}$ generated the {user's permissions}. The bit
   $\xi_{i\sep d}$ selects whether the noise-free bit $x^S_{i\sep d}$
   or the noise bit $x^N_{i\sep d}$ is observed. This model is an
   extension of the model in Fig.~\ref{fig:genModel_basic}.}
    \end{minipage}
\end{figure}

 With the structure and noise distribution combined, the resulting probability of an observed $x_{i\sep d}$ is
\begin{equation}
p\!\left(x_{i\sep d}\!\left.\right|\lset_i,\boldsymbol{\beta},
        r,\mathbf{\xi}_{i\sep d} \right) \! =\!
    p_N \!\left( x_{i\sep d}\!\left.\right|r\!\right)^{\xi_{i\sep d}}
    p_S \!\left( x_{i\sep d}\!\left.\right|\lset_i,\boldsymbol{\beta}\right)
        ^{1-\xi_{i\sep d}} \ . \label{eq:SNmixture}
\end{equation}
In Appendix~\ref{app_margXi} we marginalize out $\xi$ and obtain the final likelihood of this model.
\begin{align}
 p_{\text{MAC}}\left(\mathbf{X}\left.\right|\mathbf{Z},\boldsymbol{\beta}, r,
        \epsilon \right)
 &= \prod_{i\sep d} \left(
        \epsilon \cdot p_N \left(x_{i\sep d}\left.\right|r\right)
        + (1-\epsilon) \cdot p_{\flatRBAC}\left(x_{i\sep d} \left.\right|            \mathbf{Z},\boldsymbol{\beta} \right) \right)
 \ .  \label{mac_likelihood}
\end{align}
The generation process underlying this model is depicted in Figure~\ref{fig:genModel}.
{The advantage of an explicit noise process over threshold-based denoising methods \cite{Colantonio09:stable} is the ability to automatically adapt to the {data's} noise level. Another way to deal with noise is  denoising  in the continuous domain as a preprocessing step for a generic role mining method. Such an approach using SVD has shown good performance \cite{Molloy:2010:MRN}. While this step is computationally inexpensive, it requires selecting a cutoff threshold.}

\section{Learning the model parameters} \label{sec_optimization} 

In the last section we derived a class of probabilistic models. To apply
these models to the role mining problem, we require
algorithms that learn the model parameters from a given
access control matrix. 
In this section we {present} two such learning
algorithms. In 
particular, we use an annealed version of expectation-maximization
(EM) for the models with a fixed number of parameters and Gibbs
sampling for the non-parametric model variants.

\subsection{Annealed EM}
When applying the well-known EM algorithm to clustering problems, one
alternates between updating the expected cluster assignments given the
current centroids (E-step) and updating the centroids given the
current assignments (M-step).  In the case of the proposed model for
multi-assignment clustering (MAC), the E-step computes the expected
assignment  $\gamma_{\lset,i}$ of a user $i$ to role set $\lset$  for
each role and each user: 
\begin{equation}
\gamma_{\lset,i}
    =  p( \lset \vert \mathbf{X}, \boldsymbol{\beta} , \epsilon,r
      )^{1/T} 
  \biggr(\sum_{\lset=1}^{\left|\mathbb{L}\right|}p( \lset \vert
      \mathbf{X}, \boldsymbol{\beta} , \epsilon,r )^{1/T} \biggr)^{-1} .
    \label{eq_estepProb}
\end{equation}
Here, we have introduced the computational temperature
$T>0$. The case with $T=1$ reproduces the conventional
E-step. The limit $T\rightarrow \infty$ yields the uniform distribution over all
role sets and a low temperature $T\rightarrow 0$ makes the expectation
of the assignments ``crisp'' (close to $0$ or $1$). The normalization
ensures that the sum over all $\lset\in\mathbb{L}$ equals $1$.

While iterating this modified E-step and the conventional M-step, we decrease the temperature starting from a value of the order of the
negative log-likelihood costs. As a result, the local minima of the
cost function are less apparent in the early stage of the
optimization. In this way, lower minima can be identified than with
conventional EM, although there is no guarantee of finding the global minimum. In addition
to this robustness effect of the annealing scheme, we obtain the desired effect that, in the low-temperature regime of the optimization, the
user-role assignments are pushed towards 0 and 1. As we are ultimately
interested in binary user-role assignments, we benefit from forcing
the model to make crisp decisions.

We provide the update equations of all model parameters used in the
M-step in Appendix~\ref{ssec:Updates_generalForm}. {The rows of
  $\boldsymbol{\beta}$ are initialized with random rows of
  $1\!\!-\!\!\mathbf{X}$ and $(\epsilon,r)$ are initialized with
  $(0.1,0.5)$.} The annealed 
optimization is stopped when the last user was assigned to a single
role set with a probability exceeding $1-10^{-6}$.

\subsection{Gibbs sampling} \label{sec_GibbsSampl}
For nonparametric model variants as, for instance, the DDM with a Dirichlet process for the user-role
assignments, we employ Gibbs sampling to learn the model
parameters. Gibbs sampling iteratively samples the assignment of a
user to one of the currently existing roles or to a new role while
keeping the assignments of all other users fixed. This scheme
explicitly exploits the exchangeability property of these models. This
property states that the ordering in which new objects are randomly
added to the clusters does not affect the overall distribution over
all clusterings.

All terms involved in the sampling step are derived in
Appendix~\ref{app_eviDDM}. The probability for assigning the current
user to a particular role is given in Eq.~(\ref{samplingEq}).  The
Gibbs sampler alternates between iterating over all user-role
assignments and over all permission-role assignments. It stops if
the assignments do not change
significantly over several consecutive iterations 
or if a predefined maximum number of alternations is
reached. While running the sampler, the algorithm 
{stores} the
state with the maximum a-posteriori probability and reports this state
as the output. {This {book-keeping} leads to worse scores than computing the
  estimated score by averaging over the entire chain of sampled RBAC
  configurations. However, this restriction to a single solution
  reflects the practical constraints of the role mining
  problem. Ultimately, the administrator must choose a single RBAC
  configuration.}

\section{Experiments}\label{sec_expsFull}
In this section we experimentally investigate the proposed models on both
artificial and real-world datasets. We
start by comparing
MAC {and} 
DDM on datasets where we vary the noise level. Afterwards we compare MAC and
DDM with other methods for Boolean matrix factorization on a {collection} of
real-world datasets.

\subsection{Comparison of MAC and DDM}
\label{sec:expMACvsDDM}

MAC and DDM originate from the same core model. However, they differ in the following
respects:
First, DDM has one extra layer of roles. This additional layer,
encoded in the assignment matrix $\mathbf{Y}$, creates a clustering of
the permissions. Therefore, DDM has a two-level role hierarchy while
MAC models flat RBAC.
Second, DDM has additional constraints on its assignment matrices
$\mathbf{Z}$ and $\mathbf{Y}$ dictating that the business roles must
be disjoint in terms of their users and the technical roles must be
disjoint in terms of their permissions. 
The assignments of
business roles to technical roles  $\mathbf{V}$  are unconstrained. MAC
has no constraints at all.
A user can have multiple roles and permissions can be assigned to
multiple roles.
The last difference between the two models are the prior assumptions on the
model parameters. While MAC implicitly assumes uniform prior
probabilities for its parameters, DDM makes explicit non-uniform
assumptions encoded in the Beta priors and the Dirichlet priors.

To evaluate which of the two model variants is best suited to solve the
role mining problem, we design the following experiment. We generate
access control data in two different ways. One half of the datasets is
generated according to the MAC model. We take a set of roles and
randomly assign users to role combinations to create a user-permission
assignment matrix. Then, we randomly select a defined fraction of
matrix entries and replace them with the outcome of a fair coin flip.
Some users in these datasets are generated by multiple {roles}. The
second set of datasets is generated from the DDM probability
distribution by repeatedly sampling business roles and technical roles
from the Dirichlet process priors and randomly connecting them
according to the Beta-Bernoulli probabilities.

On both kinds of data sets, we infer RBAC configurations with DDM and
with MAC. In this way, the model assumptions always hold for
one of the models while the other one operates in a model-mismatch
situation.  Moreover, data from DDM can have an arbitrary number of
underlying roles. Therefore, MAC requires an additional model-order
selection mechanism.  Cross-validation with the generalization test
described in Section~\ref{rmDef_qualmeas} is employed as a quality
{measure}.

We control the difficulty of the experiments by varying the noise
level in the data. For each noise level, we sample 30 datasets  from
each model variant{, each with 400 users and 50 permissions}. On each dataset, we run each model 10 times and
select the solution with the highest internal quality score,
respectively.  Finally, we evaluate the inferred roles based on their
generalization error on a hold-out test set.

\begin{figure}
\centering
\includegraphics[width=0.49\textwidth]{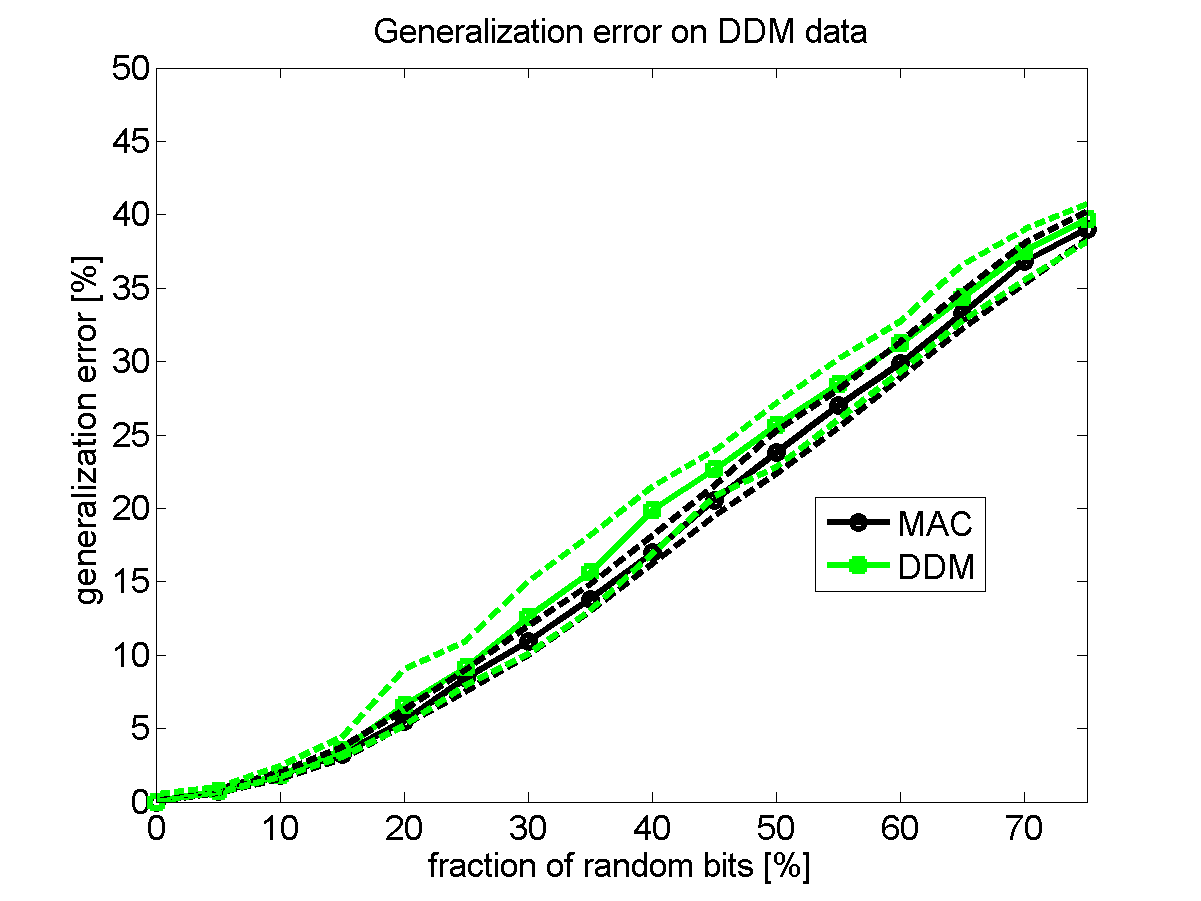}
\includegraphics[width=0.49\textwidth]{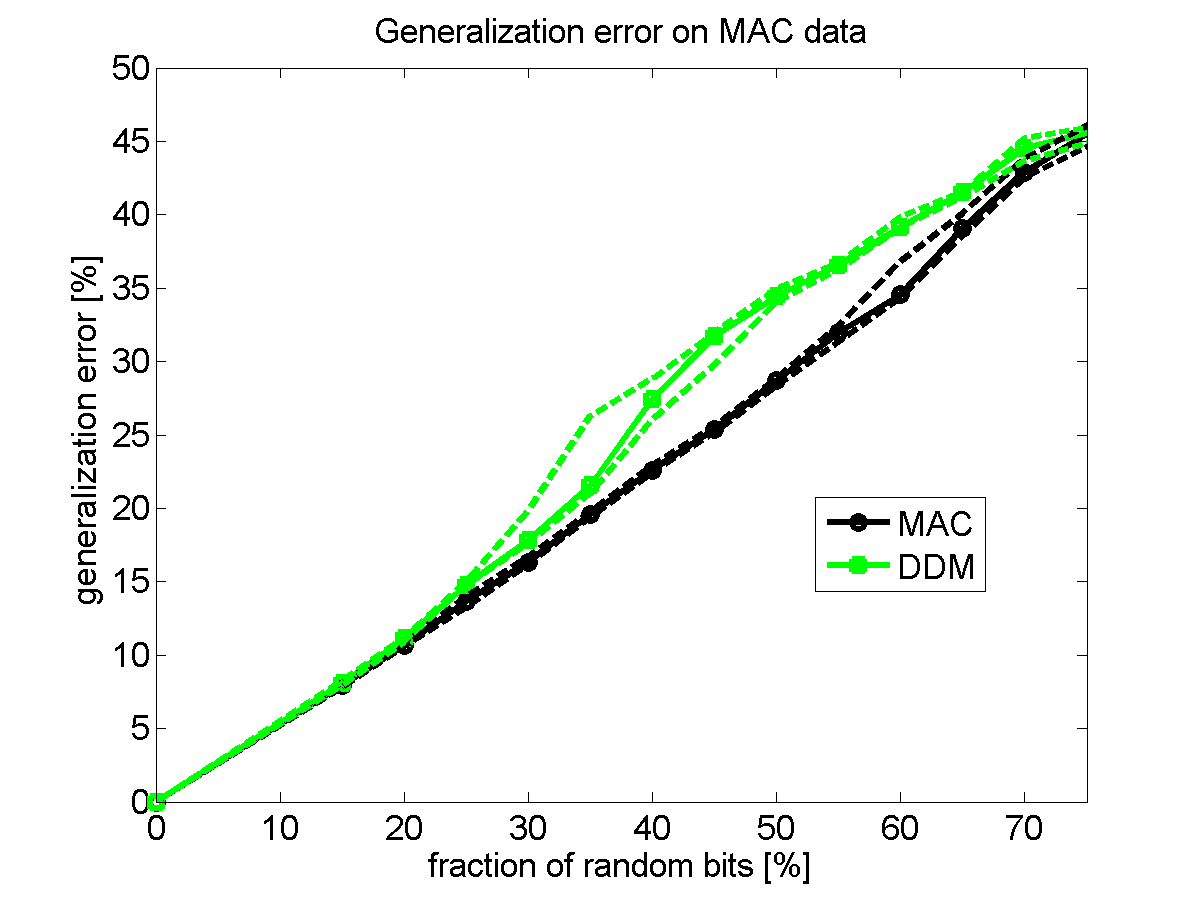}
\caption{Comparison of the generalization error of MAC and DDM on data
  generated from a DDM role hierarchy and a flat RBAC configuration
  respectively. 
  \label{fig_IRMvsMAC}}
\end{figure}

\paragraph{Results}
We report the median generalization error of the inferred matrix
decompositions and the 25\%/75\%-percentiles in
Figure~\ref{fig_IRMvsMAC}. The left plot depicts the generalization
error on DDM data and the right plot {shows} the error on MAC data.

{We see} that the overall trend of both models
is similar for both types of data. The generalization error increases
with increasing noise. {There are two explanations for this
  behavior.}  First, the problem of
estimating the structure of the data becomes increasingly difficult
 when increasing the noise level. Second, noisy bits are likely to
be wrongly predicted, even when the data structure is learned well.

{We also see} that MAC and DDM generalize almost equally
well on DMM data. MAC is even slightly better than DDM. In contrast,
DMM achieves a worse generalization error than MAC on MAC data in the
intermediate noise range. One would expect that each model generalizes
best on data that is generated according to its
assumptions. This {behavior} can be observed on MAC data. However,
on DDM data MAC is as good as DDM.
The reason is that for DMM data, the model assumptions of MAC are in
fact not violated.  Even though DMM has an extra layer of roles, this
model instance is less complex than flat RBAC (which MAC models). One
can see this by collapsing one DDM layer. For instance, define
$\mathbf{u'}:=\mathbf{v}\BMP \mathbf{y}$. Then permissions can be
assigned to multiple roles (because there are no constraints on the
business-role to technical-role assignment $\mathbf{v}$). At the same
time, $\mathbf{z}$ still provides a disjoint clustering of the
users. In this flat 
RBAC configuration $(\mathbf{z},\mathbf{u'})$ the roles overlap
in terms of their permissions but not in terms of the users that are
assigned to the roles. The same 
model structure arises in single-assignment clustering (SAC), a constrained
variant of MAC, where users can have only one role.  As a
consequence, we can interpret DDM as a SAC model with prior
probabilities on the model parameters.  In contrast to MAC,
SAC yields inferior parameter
estimates as it must fit more model parameters for the
same complexity of data. Hence, it has a larger
generalization error.

{In contrast to the structural difference between DDM and
MAC, the differences in the optimization algorithm and in the
Bayesian priors  have only a minor influence on the results.} It appears that
MAC can compensate for a missing internal mechanism for finding
the appropriate number of roles when an external validation step is
provided. Also, the Gibbs sampling scheme and the deterministic
annealing algorithm perform equally well on the DDM data. Given that
the prior Beta distributions for the Bernoulli variables of the DDM
provide no improvement over MAC, it seems unnecessary to extend MAC with such priors.

We experimentally confirmed that MAC is a more general model than
DDM. {We also found that} the generalization error provides a good criterion for selecting
 the number of roles without making explicit prior assumptions
on the distribution of this number. We therefore recommend using the MAC model for role mining with real-world 
datasets. 

{
\subsection{Noise}
\label{sec:expNoise}
In this section we focus on erroneous user permission assignments. As
explained in Section~\ref{rmDef_newDef}, the user-permission
assignment matrix $\upa$ given as an input {to} the role mining problem
contains exceptional or 
{erroneous} entries, which we refer to as noise. More precisely,
we assume that a hidden noiseless assignment 
$\upa'$ exists, but only $\upa$, a noisy version  of it, is
observable. By inferring a role structure $\rc=
(\mathbf{z},\mathbf{u})$ that supposedly underlies the input matrix
$\upa$, our probabilistic models {approximately reconstruct} the noiseless
matrix as $\mathbf{\hat{z}}\BMP\mathbf{\hat{u}}$.  
In synthetic experiments, one can compare this reconstruction against the noise-free assignment $\upa'$. 
We investigate two questions. First, how conservative are the RBAC
configurations that our algorithms find, that is,  how many new errors are introduced in the reconstruction? Second, does the probabilistic approach provide a measure of confidence for reconstructed user-permission assignments?

Figure~\ref{fig_errorAnalysis} depicts error rates that provide answers to both questions. In  Figure~\ref{fig_errOldvsNew} we contrast the fraction of wrong reconstruction assignments that are new with the fraction of wrong input assignments that have not been discovered. To obtain these values, we ran experiments on input matrices created from the DDM with a fraction of randomized bits that varies from 5\% to 75\%. As can be seen, MAC's ratio of newly introduced to repeated errors is constant for false negatives and false positives alike. DDM tends towards repeating old errors and introduces fewer new errors, while the sum of new and old errors is approximately the same as for MAC. The maximal sum of error rates is 8\% false negatives and 7.5\% false positives, which is small compared to the maximal fraction of random bits 75\%.

\begin{figure}[htb]
\centering
\subfloat[New vs. old errors]  {
\centering
\includegraphics[width=0.39\textwidth]{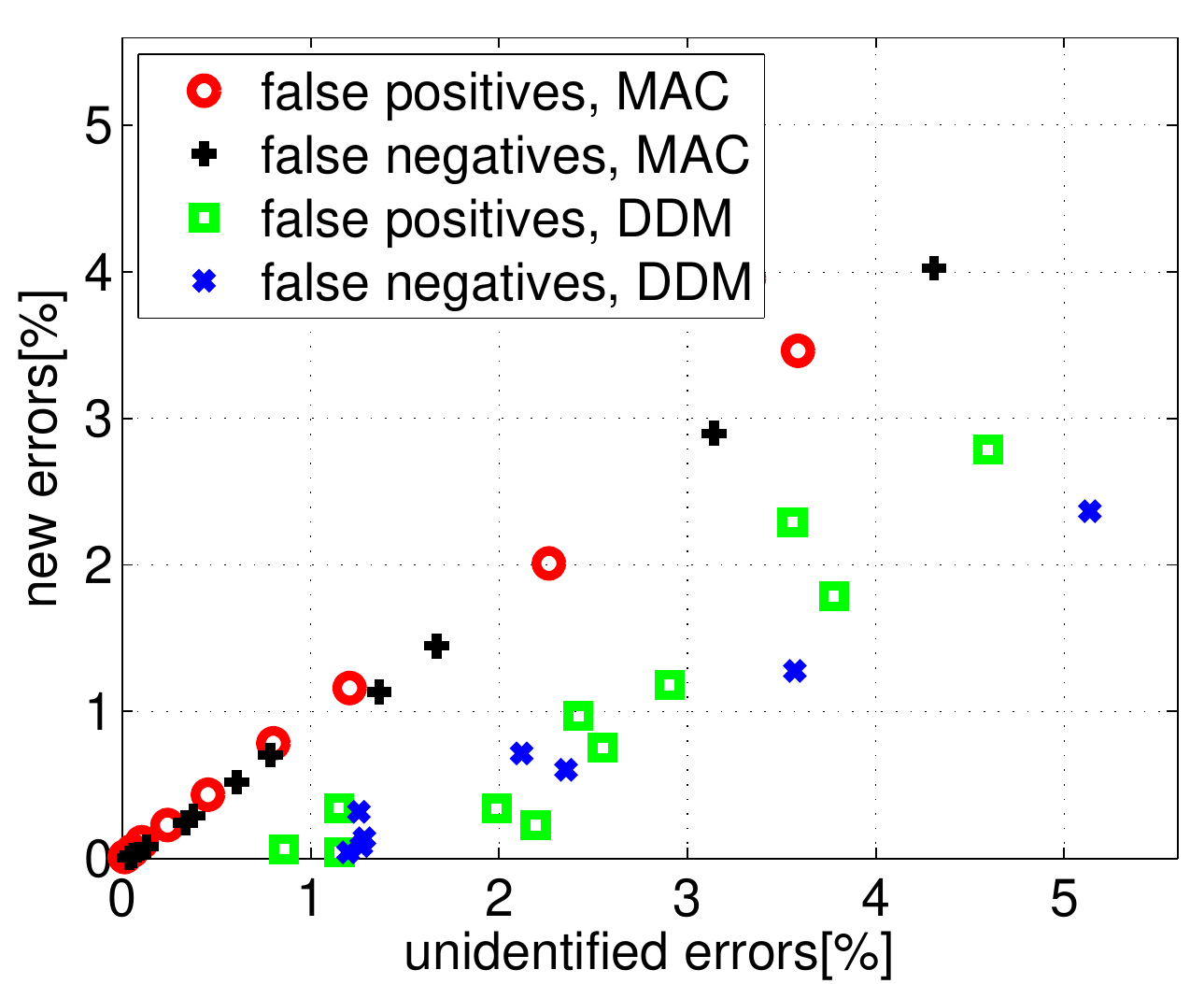}
\label{fig_errOldvsNew}
}
\ \ \ \ \ \ \ \ \ 
\subfloat[Assignment Confidence]{
\centering
\includegraphics[width=0.4\textwidth]{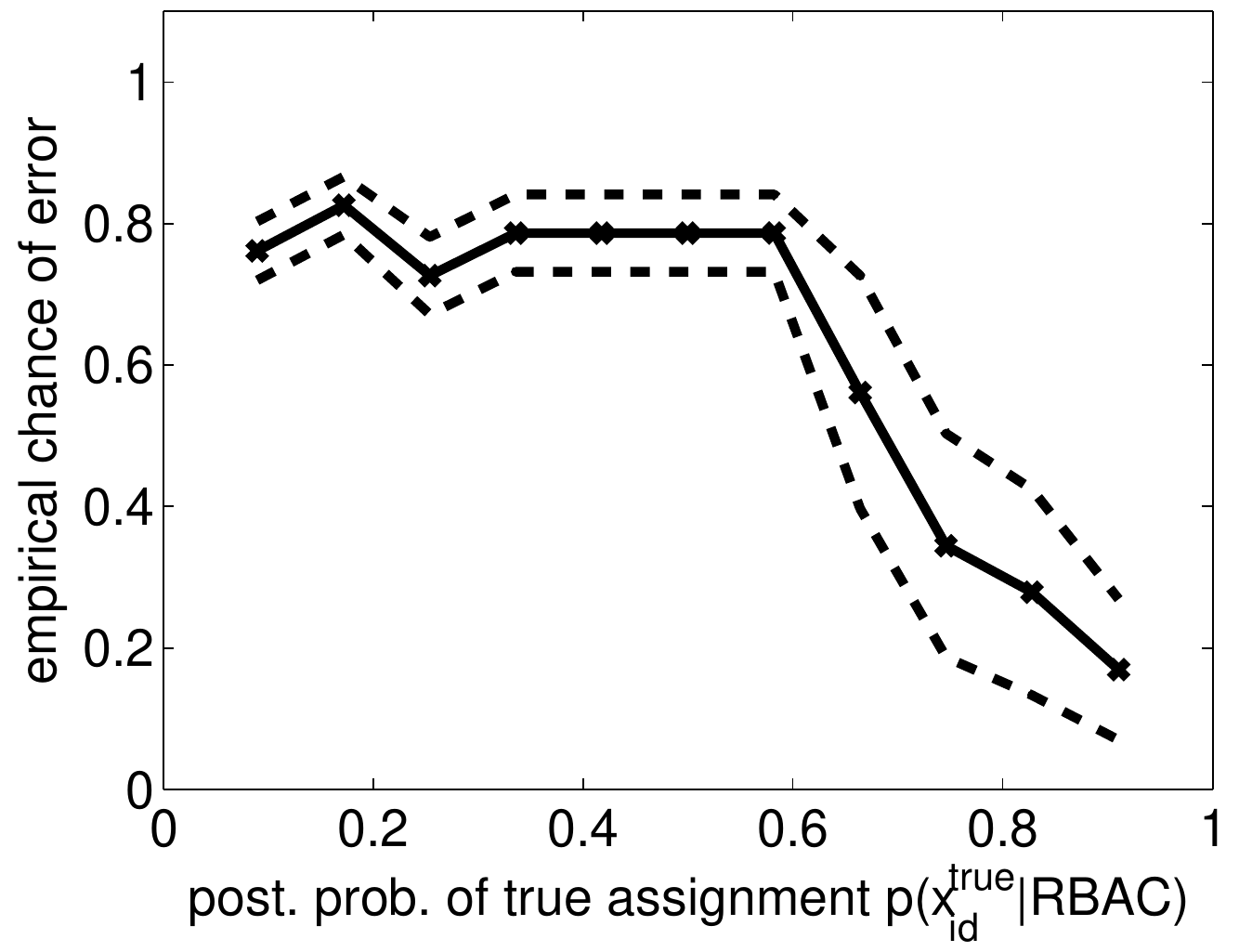}
\label{fig_errVSposterior}
}
\caption{{Error analysis. \ref{fig_errOldvsNew}: How many new
    errors are added to the assignment versus how many original errors were
    not identified? We distinguish false negatives from
    false positives.  \ref{fig_errVSposterior}: Trend of the empirical
    chance of creating a wrong assignment $x_{id}$ as a function of
    the posterior probability of the correct assignment given the
    learned RBAC configuration. It is apparent that the posterior
    provides a measure of confidence.} 
  \label{fig_errorAnalysis}}
\vspace{-2em}
\end{figure}

Figure~\ref{fig_errVSposterior} illustrates the empirical probability
that a reconstructed user-permission assignment is wrong. The x-axis
is the posterior probability of the true value $p(x_{id}'\vert\rc)$ of
this assignment, i.e. the probability of reconstructing the assignment
correctly as computed by the learned model. The plot 
{convincingly} 
illustrates that the posterior provides a measure of confidence. The
stronger the model prefers a particular assignment (the closer the
posterior is to 1), the less likely it is to introduce an error if one
follows this preference. This {model property} means that, in addition to the RBAC
configuration, our algorithms can output a confidence score for each
resulting user-permission assignment. This {uncertainty estimate}
could  help practitioners when configuring RBAC using our methods.  

}

\subsection{Experiments on real-world data} \label{sec_RWexps} In this
section we compare MAC and DDM with other Boolean matrix
factorization techniques on real-world datasets.

The first dataset {\verb|LE-access|} comes from a Large Enterprise.
It consists of the user-permission assignment matrix of $N=4900$ users and
$D=1300$ permissions as well as business attributes for each user. In
this section, we will ignore the business attributes but in
Section~\ref{hybrid_sec:risk} we will include them in a hybrid role
mining process.  The other datasets are six publicly available
datasets from HP~labs \cite{EneEtAl}. They come from different systems
and scenarios. The dataset {\verb|dominos|} is
an access-control matrix from a Lotus Domino server, 
{\verb|customer|} is the access-control matrix of an HP customer,
{\verb|americas small|} and {\verb|emea|} are access-control
configurations from Cisco firewalls, and {\verb|firewall1|} and
{\verb|firewall2|} are created from policies of Checkpoint firewalls.

We compare several algorithms for Boolean matrix factorization on
these datasets. In addition to DDM and MAC, there are two other
probabilistic methods that have been developed in different
contexts. Binary Independent Component Analysis (BICA) \cite{BICA}
learns binary vectors that can be combined to 
fit the data. These
vectors, representing the roles in our setting, are orthogonal,
that is, each permission can be assigned to only one role. In
\cite{IBPbinary}, an Indian buffet process has been combined with a
noisy-OR likelihood to a nonparametric model that learns a Boolean
matrix factorization. We call this model infinite noisy-OR (INO). The
noisy-OR likelihood is closely related to the likelihood of MAC. The
difference is that its noisy bits are always flipped to 1, whereas
in MAC a noisy bit is a random variable that is 1 with probability
$r$ and 0 otherwise. Similar to the Dirichlet process used in DDM,
the Indian buffet process in INO is capable of learning the number of
factors (here the number of roles). Another method that we compare
with is a greedy combinatorial optimization of a matrix factorization
cost function. This method, called Discrete Basis Problem solver
(DBPS), was proposed in \cite{Miettinen}. It finds a Boolean matrix
decomposition that minimizes the distance between the input matrix and
the Boolean matrix product of the two decomposition matrices. This
distance weights false 1s {differently compared to} false
0s, with weighting 
factors that must be selected. The decomposition is successively
created by computing a large set of candidate vectors (here candidate
roles) and then greedily selecting one after the other such that in
each step the distance function is minimized.

For each dataset, we randomly subsample a training set containing 80\%
of the users and a hold-out test set containing the remaining 20\%
users. All model parameters are trained on the training set and the
generalization error is evaluated on the test set. We repeat this
procedure five times with a different random partitioning of the
training set and the test set.

We train the model parameters as follows. DDM and INO select
the number of roles internally via the Dirichlet process and the
Indian Buffet process, respectively. For MAC, BICA, and DBPS, we
repeatedly split the training data into random subsets and compute the
{validation} error. We then select the number of roles (and other
parameters for BICA and DBPS, such as thresholds and weighting
factors) with lowest {validation} error and train again on the
entire training set using this number.

\begin{table}[htb]
\vspace{-1em}
\caption{Results on HP~labs data \label{tab:real_results}}{
\begin{tabular}{|l||r|r|r||r|r|r|}
\hline
  & \multicolumn{3}{|c||}{{ customer} }
   &\multicolumn{3}{|c|}{{ americas small} }
    \\
   &\multicolumn{3}{|c||}{10,021 users $\times$ 277 perms. }
   & \multicolumn{3}{|c|}{3,477 users $\times$ 1,587 perms. }\\
\hline
  & $k$ & gen. error [\%] & run-time [min]
  & $k$ & gen. error [\%] & run-time [min]
  \\
\hline
  MAC  & 187 & $ {{2.40 \pm 0.03}} $     
          & {49}  & 139 & {$ 1.03 \pm 0.01 $} & {80}          \\
          DDM & 4.6 &  {$1.90 \pm 0.01 $}& {800}
                 & 48.8 & {$ 1.00 \pm 0.03 $} & {2000} \\
  DBPS & 178 & {$ 2.54 \pm 0.05 $} & {43}
         & 105 & {$ {1.00 \pm 0.03} $} & {100} \\
  INO  & 20 & {$ 7.8\phantom{0} \pm 1.6\phantom{0} $} &{ 990}
         & 65.6 & {$ 1.05 \pm 0.01 $} &{ 300}\\
  BICA
       & 82 & {$ 2.66 \pm 0.02 $} & {200}
       & 63 & {$ {1.00 \pm 0.01}$} &{ 64}\\
\hline
\multicolumn{7}{c}{ }    \\
\hline
  & \multicolumn{3}{|c||}{{ firewall1}}
  & \multicolumn{3}{|c|}{{ firewall2}} \\
  & \multicolumn{3}{|c||}{365 users $\times$ 709 perms.}
  & \multicolumn{3}{|c|}{325 users $\times$ 590 perms.}
  \\
\hline
  & $k$ & gen. error [\%] & run-time [min]
  & $k$ & gen. error [\%] & run-time [min]\\
\hline
  MAC  & 49 & {$ {4.57 \pm 0.01} $} & {10\phantom{.0}}
       & 10 &{ $ {3.40 \pm 0.00} $}& {1.8} \\
       DDM 
       & 24 & {$ 4.52 \pm 0.01 $}& {38\phantom{.0}}
       & 9.6 & {$ 11.28 \pm 0.00 $} & {5.4}\\
  DBPS & 21 &  {$13.6\phantom{0} \pm 3.1\phantom{0} $}&  {5\phantom{.0}}
       & 4 & { $ 19.5\phantom{0} \pm 4.4\phantom{0} $} &  {2\phantom{.0}} \\
  INO  & 38.2 & {$  8.04 \pm 0.00 $}&  {96\phantom{.0}}
       & 6.2 & { $ 11.15 \pm 0.00 $}&{ 14\phantom{.0}}\\
  BICA & 18 &{ $ 12.8\phantom{0} \pm 3.0\phantom{0} $}&{ 2.1}
       & 4 & {$ 19.9\phantom{0} \pm 4.5\phantom{0} $} & {0.9} \\
\hline
\multicolumn{7}{c}{ }    \\
\hline
   &\multicolumn{3}{|c||}{{ dominos } }
   &\multicolumn{3}{|c|}{{ emea} }
    \\
   &\multicolumn{3}{|c||}{79 users $\times$ 231 perms. }
   & \multicolumn{3}{|c|}{35 users $\times$ 3,046 perms. }\\
\hline
  & $k$ & gen. error [\%] & run-time [min]
  & $k$ & gen. error [\%] & run-time [min]
  \\
\hline
MAC   & 7 & {$ 1.73 \pm 0.00 $}& {1.1}
          & 3 &{ $ 8.7\phantom{0} \pm 1.2\phantom{0} $}&  {0.7}
          \\
   DDM & 6.4 &{ $ 1.70 \pm 0.01 $} &  {2.4}
          & 18.2 &{ $ 8.4\phantom{0} \pm 0.1\phantom{0} $} & { 143\phantom{.0}}\\
  DBPS & 9 &{ $ 2.3\phantom{0} \pm 0.5\phantom{0} $}& { 0.2}
         & 8 & {$ {7.3\phantom{0} \pm 2.6\phantom{0}} $}& { 1.1} \\
  INO  & 26 &{ $ {1.7\phantom{0} \pm 0.1\phantom{0}} $}& { 9.0}
         & 80.4 & {$ 10.1\phantom{0} \pm 2.4\phantom{0} $} & { 204\phantom{.0}}\\
  BICA
       & 3 & {$ 1.9\phantom{0} \pm 0.3\phantom{0} $} &  {0.1}
       & 5 & {$ 8.6\phantom{0} \pm 2.8\phantom{0} $}& { 1.0}\\
\hline
\end{tabular}}
\vspace{-0.5em}
\end{table}

\begin{figure}[tb]
\centering  \includegraphics[width=0.5\columnwidth]{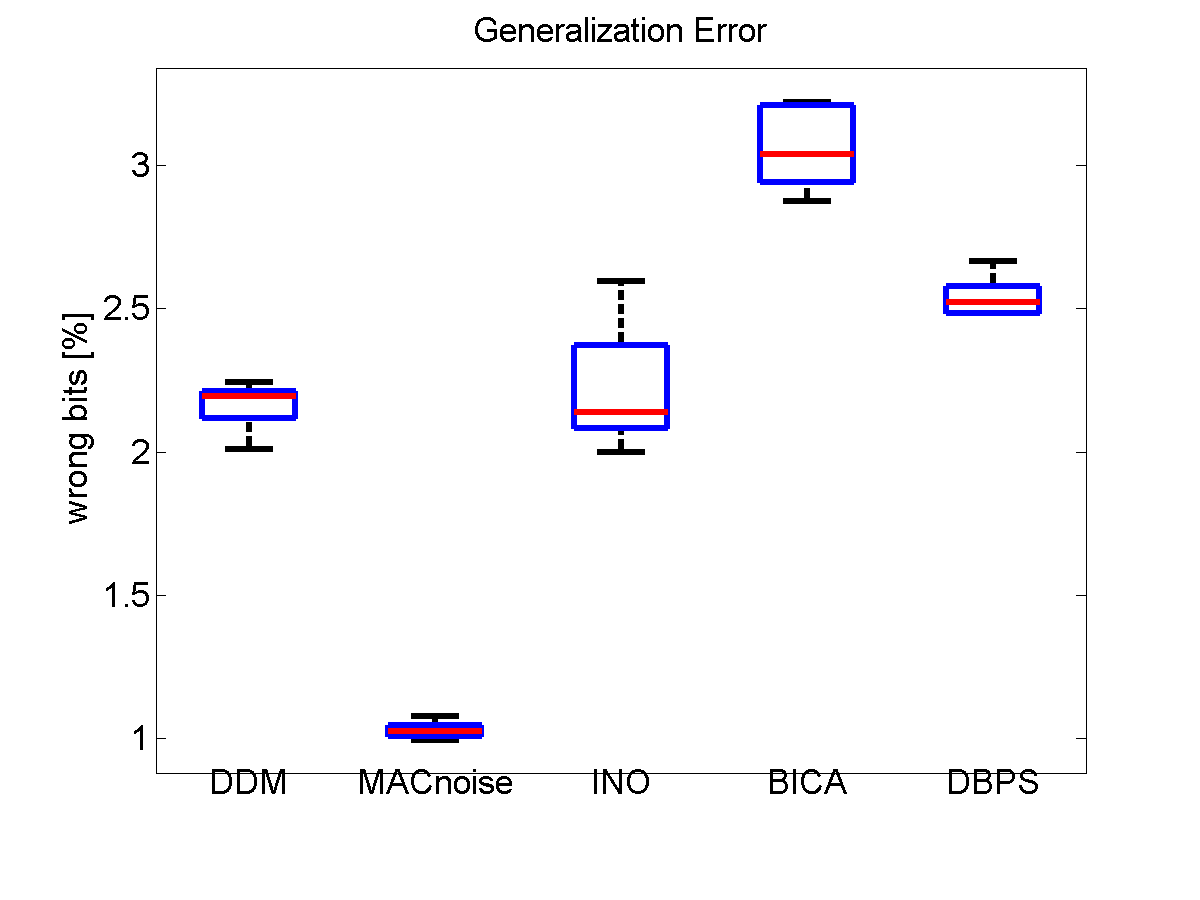}  
\caption{ \label{fig:genError_original} Generalization error on real-world data.    }
\end{figure}

Our experimental findings on the HP datasets are summarized in
Table~\ref{fig:genError_original}. 
We report the
number of users and permissions and scores for each
method: the number of roles {discovered}, the median generalization error and
its average difference to the 25\% and 75\%-percentiles, and the
{the time required for} one run.  For {\verb|americas small|} and {\verb|emea|}
the {errors of all methods are} within %
the percentiles. For {\verb|customer|}, DDM
generalizes best. For {\verb|firewall1|} and {\verb|dominos|}, DDM
and MAC perform equally well and lead the ranking, whereas for
{\verb|firewall2|} DDM is inferior to MAC. {While for all datasets MAC finds solutions with an error close to the solutions of the ``winner'', DDM deviates largely from the best method in the case of} {\verb|firewall2|}.

All methods differ significantly in runtime with DBPS always being
the fastest method. However, the run-times are difficult to compare
{for several reasons. First}, all algorithms have been implemented by
different authors in different programming languages. {Second, they
  all use different stopping criteria. We manually selected these   criteria  such that each method achieves good results on training data, but it is impossible to tune them for a fair runtime
  comparison. Finally, all algorithms, except for INO and DDM, must be
  run several times to search for appropriate parameters.
  {The runtimes
  reported in Table~\ref{tab:real_results} account for one such run.
  To find the appropriate $k$ for MAC, DBPs, and BICA, we start with $k=2$ and increase $k$ until the generalization error significantly increases. The final value of $k$ given in Table~\ref{tab:real_results} is thus indicative of the number of runs. For DBPs, we additionally tuned the parameters ``threshold'' and ``bonuses'' with a grid search over 25 candidate values.}}

\medskip The results on {\verb|LE-access|} 
are graphically illustrated in Figure~\ref{fig:genError_original}. The
MAC model generalizes best.  The two nonparametric
Bayesian models DDM and INO have a similar performance. DBPS performs
a bit worse and BICA has the highest generalization error.
Apparently, on this dataset, the assumption of orthogonal roles does
not hold.

\section{Hybrid role mining}
\label{hybrid_sec:risk} 
Hybrid role mining accepts as input additional information 
{on} the attributes of the users or business processes. The goal, as defined by
the inference role mining problem, remains unchanged: Find the RBAC
configuration that most likely underlies the observed user-permission
assignment. As this configuration is assumed to reflect the business
properties and the security policies of the enterprise, we approach
the hybrid role mining problem by jointly fitting an RBAC
configuration to the user-permission assignment and to the business
information given as additional input.

We account for this additional information by modifying the
optimization problem for role mining. {The original problem is to {minimize the negative log-likelihood for the MAC model (\ref{mac_likelihood})}.} The original cost
function is
\begin{align}
R^{(ll)}    &=\sum_{i,\lset}z_{i\lset} R_{i,\lset}^{(ll)}\  , \  \
z_{i\lset}\in \{ 0,1\}, \  \  \forall i:\sum_{\lset} z_{i\lset}=1 \, .
\label{hybrid_eq_risk_full}
\end{align}
Here we used the assignments $z_{i\lset}$ from user $i$ to the set of
roles $\lset$. With these assignments and with the assignments
$z_{\lset k}$ from role sets to roles, a user is assigned the roles,
which are contained in the role set that he is assigned to
($
z_{ik}=\sum_{\lset}z_{i\lset}z_{\lset k} 
$). 
The individual costs of assigning a user $i$ to the set of roles $\lset$ is
\begin{align}
R_{i,\lset}^{(ll)}
    &= -\log\left( \prod_d
        p_M\left(x_{id}\left.\right|\lset,\mathbf{\beta}, r,
        \epsilon \right) \right)  \, .
\label{hybrid_eq_risk}
\end{align}

{We now add an  additional term to this negative log-likelihood cost function to define an objective function for hybrid role mining. We use a linear}
combination of the likelihood costs Eq.~(\ref{hybrid_eq_risk_full})
and a term $R^{(S)}$ that accounts for business information costs:
\begin{equation}
R :=   R^{(ll)}\!/D +\lambda R^{(S)}\, ,
    \label{hybrid_risk_hybrid}
\end{equation}
where $\lambda \geq 0$ 
weights the influence of the business information. The term $1/D$
makes the likelihood costs independent of the number of permissions 
$D$.

Role mining without business information is a special case of
Eq~(\ref{hybrid_risk_hybrid}), where $\lambda=0$, whereby role mining
optimizes the model parameters with respect to
Eq.~(\ref{hybrid_eq_risk_full}).  This problem has a huge solution
space, spanned by the model parameters, with many local minima. By
incorporating business information into this optimization problem
($\lambda>0$), we introduce a bias toward minima that correspond 
to the business attributes used as additional input.

We consider an RBAC configuration to be meaningful if employees
satisfying identical business predicates (that is, having the same business
attributes) are also assigned to similar (ideally identical) sets of
roles.  To account for this, we propose an objective function $R^{(S)}$
that computes the similarity of the role assignments of all pairs of
users with the same business attributes.  $R^{(S)}$ compares all pairs
of users $(i,i')$ having the business attribute $s$ with respect to
their role assignments $(z_{i\cdot}$, $z_{i'\cdot})$. Let
$w_{is}\!\in\!\{0,1\}$ encode whether user $i$ has business attribute
$s$ ($w_{is}=1$) or not ($w_{is}=0$). Then the cost of a user-role
assignment matrix $\mathbf{Z}$ is
\begin{eqnarray}
 R^{(S)} \!\! &=& \!\! \frac{1}{N} \sum_s
         \sum_{i,i'} w_{is}w_{i's}
         \sum_k z_{i'k}\left( 1-2z_{i'k}z_{ik}\right) .
        \label{hybrid_eq_pairwCosts}
\end{eqnarray}
$N$ is the total number of users and $k\in \{1, \ldots,K\}$ is the role
index. Each user has a single business attribute $s$,  that is $\sum_s
w_{is}=1$, but can be assigned to multiple roles, $1 \leq \sum_k
z_{ik}\leq K$. The term $ \sum_k z_{i'k}\left(
  1-2z_{i'k}z_{ik}\right)$ in Eq.~(\ref{hybrid_eq_pairwCosts})
computes the agreement between the binary assignment vectors
$(z_{i\cdot},z_{i'\cdot})$ for all pairs of users $(i,i')$ having the
same attribute $s$ (which is the case iff $w_{is}w_{i's}\!=\!1$). The
subterm $1-2z_{i'k}z_{ik}$ switches the sign of a single
term such that agreements ($z_{ik}z_{i'k}=1$) are rewarded and
differences ($z_{ik}z_{i'k}=0$) are penalized.  An alternative to
Eq.~(\ref{hybrid_eq_pairwCosts}) would be to compute the Hamming
distance between the two assignment vectors. However, this  has
the drawback of penalizing pairs with differently sized role sets. We
have chosen the dissimilarity function in
Eq.~(\ref{hybrid_eq_pairwCosts}) to avoid a bias towards equally
sized role sets. 

{Note that our objective function conceptually differs from those proposed in \cite{italians2,molloy,Molloy:multiObjectives,xu12algorithms}. While we minimize 
the role dispersion of users with the same attributes, their objective
is to minimize measures of attribute dispersion \cite{xu12algorithms} of users with the
same roles. We see two advantages of our cost function. First, it
enables  multiple groups of users with different attributes to share
roles, such as users of all organizational units getting a ``standard
role''. Second, it is easier to add new users in our framework as
usually the attributes of new users are given whereas the roles are
not.}

\subsubsection*{Optimization} \label{hybrid_sec:inference} 
We now demonstrate how to optimize the new cost function for hybrid
role mining using the deterministic annealing framework presented in
Section~\ref{sec_optimization}.  Specifically, we convert the term
$R^{(S)}$ into a form that enables us to conveniently compute the
responsibilities $\gamma_{i\lset}$ in the E-step.

The responsibility $\gamma_{i\lset}$ of the assignment-set  $\lset$ for data item $i$ is given by
\begin{equation}
\gamma_{i\lset}
    := \frac{\exp\left( -R_{i\sep \lset} /T\right)}
         {\sum_{\lset'\in\lsetset}
            \exp\left( -R_{i\sep \lset'} / T\right)}\ .
    \label{eq:gammaiL_generalForm}
\end{equation}
In Eq.~(\ref{eq_estepProb}), the individual cost terms were $R_{i\sep
  \lset}=-\log\left( p(\mathbf{x_{\allInd i}} \vert
  \lset_i,\boldsymbol{\beta}, \epsilon , r )\right) $. Now the full
costs are extended with $R^{(S)}$.  To compute
Eq.~(\ref{eq:gammaiL_generalForm}), we shall rewrite $R^{(S)}$ as a
sum over the individual contributions of the users.

Let $N_{sk} := \sum_i z_{ik}w_{is}$ be the number of users that have
the business attribute $s$ and are assigned to role $k$, and let $s_i$
be the attribute of user $i$.  We first simplify and reorganize
Eq.~(\ref{hybrid_eq_pairwCosts}) using these auxiliary variables
\begin{align}
 R^{(S)} 
 &= \frac{1}{N} \sum_{s}\sum_{i}  w_{is} \sum_k  \left(N_{sk}- 2z_{ik} N_{sk} \right) \nonumber \\
 &= \frac{1}{N} \sum_{i}  \sum_k  \left(N_{s_ik}- 2z_{ik} N_{s_ik} \right)  \nonumber \\
 &= \sum_{i,k}\left( 1- z_{ik}\right)\frac{N_{s_ik}}{N}-\sum_{i,k}  z_{ik} \frac{N_{s_ik}}{N} \ . \label{hybrid_TD_risk_rolewise}
\end{align}
In this formulation, it becomes apparent that user $i$ with attribute
$s_i$ should be assigned to the role $k$ ($z_{ik}=1$) if many users have attribute $s_i$ and role $k$.

Finally, we decompose the costs into individual contributions
of users and role sets.  We use the notion of user-to-role set
assignments $z_{i\lset}$ to substitute the user-role assignments by
$z_{ik}=\sum_{\lset}z_{i\lset}z_{\lset k} $:
\begin{align}
\label{hybrid_eq_Rtd_asSumOverLsets}
R^{(S)}&=\sum_{i,k} \left( \left(1-\sum_{\lset}z_{i\lset}z_{\lset
      k}\right)\frac{N_{s_ik}}{N}-  \sum_{\lset}z_{i\lset}z_{\lset k}
  \frac{N_{s_ik}}{N} \right)  \nonumber \\ 
&=\sum_{i,\lset}z_{i\lset}
\left(\sum_{k\notin\lset}\frac{N_{s_ik}}{N}- \sum_{k\in\lset}
  \frac{N_{s_ik}}{N} \right) 
 \ \ 
 =\sum_{i,\lset}z_{i\lset} R_{i,\lset}^{(S)}\ \ . 
\end{align}
In this form we can directly compare the business cost function with
the likelihood costs given by Eq.~(\ref{hybrid_eq_risk}). We can
therefore easily compute the expectation in the E-step by substituting
the costs in Eq.~( \ref{eq:gammaiL_generalForm}) with $R_{i\sep
  \lset}= R_{i,\lset}^{(ll)} + R_{i,\lset}^{(S)}$.

In the iterative update procedure of the deterministic annealing
scheme, one faces a computational problem arising from a recursion. The
optimal assignments $z_{i\lset}$ depend on the $N_{s_ik}$, which are,
in turn, computed from the $z_{i\lset}$ themselves. To make this
computation at step $t$ of our algorithm feasible, we use the expected
assignments $\gamma_{i\lset}^{(t-1)}:=\expect{z_{i\lset}^{(t-1)}}$ of
the previous step instead of the Boolean $z_{i\lset}^{(t)}$ to
approximate $N_{s_ik}^{(t)}$ by its current expectation with respect
to the Gibbs distribution:
\begin{equation}
N_{s_ik}^{(t)} \approx  \expect{N_{s_ik}^{(t-1)}}
  =  \sum_{\lset}z_{\lset k}\sum_{i'}w_{i's_i}\gamma_{i'\lset}^{(t-1)}  \, .
\end{equation}
{We do not have a proof of convergence for this algorithm. However we observe that when running it multiple times with random initializations, it repeatedly finds the same solution with low costs in terms of business information and model likelihood.
}

{
\subsection{Selecting relevant user attributes}
}
\label{hybrid_sec:sIanalysis} 
An important step for hybrid role mining is selecting of the set of user
attributes used as input for the optimizer.  
User attributes that do not provide information about the users' permissions
should not be used for hybrid role mining. In fact,
requiring that the roles group together users with irrelevant attributes 
can result in inferred RBAC configurations
that are worse than those inferred without using the attributes.

To select appropriate user attributes, we propose an
information theoretic measure of relevance. Let the random variable
$X_d\in\{0,1\}$ be the assignment of permission $d$ to a generic user.
$S$ is the random variable that corresponds to the business attribute of
a generic user (e.g.~``job code'') and let $s$ be one of the actual
values that $S$ can take (e.g.~``accountant''). Let $p(x_d):=1/N \cdot
\sum_i x_{id}$ be the empirical probability of $d$ being assigned to an
unspecified user, and let $p(x_d|S=s):=1/N\cdot \sum_i x_{id} w_{is}$ be
the empirical probability of $d$ being assigned to a user with business
attribute $s$. With these quantities, we define the binary entropy, the
conditional entropy, and the mutual information as:
\begin{eqnarray}
h(X_d)&:=& - \!\! \sum_{x_d\in\left\{0,1\right\}} \!\!
p(x_d)\log_2\left(p(x_d)\right)\ , 
\\
h(X_d|S)&:=& \! - \!\! \sum_{s\in S}
p(s) \!\!\!\!\! \sum_{x_d\in\left\{0,1\right\}}\!\!\!\!\!
p(x_d|S=s)\log_2\left(p(x_d|S=s)\right)\ , 
\\
I(X_d;S)&:=&h(X_d)-h(X_d|S)\ .
\end{eqnarray}
The entropy $h(X_d)$ quantifies the uncertainty about whether a 
user is assigned permission $d$. The conditional entropy $h(X_d|S)$ is the uncertainty for a
user whose business attribute $S$ is known. The mutual information
$I(X_d;S)$ is the absolute increment of information about the
user-permission assignment gained by knowledge of $S$. This
number indicates how relevant the attribute $S$ is for the assignment of
permissions. There is one pitfall, though. If one compares this score on
different permissions $d$ with different entropies $h(X_d)$, then
low-entropic permissions will have a smaller score simply because there is little entropy to reduce. We therefore compute the relative mutual
information (\cite{cover_joy}, p.~45) as a measure of relevance:
\begin{equation}
\rho_d(S) :=  I(X_d;S)/h(X_d)  =
    1- h(X_d|S)/h(X_d)  \, .
    \label{hybrid_definition_rel}
\end{equation}
We use the convention $0/0:=1$ for the case where $h(X_d)=0$ (then $I(X_d;S)$ will also be $0$).
This number can be interpreted as the fraction of all bits in $X_d$
that are shared with $S$. Alternatively, $\rho_d(S)$ can be read as
the fraction of missing information about permission $d$ that is removed by the
knowledge of $S$.

\paragraph{Limit of few observations per business attribute}
One should take care to use sufficiently many observations when
estimating the relevance $\rho_d(S)$ of a business attribute $S$. With
too few observations, this measure is biased towards high
relevance. Imagine the problem of estimating the entropy of a fair coin
based on only one observation (being heads without loss of
generality). Na\"{\i}vely computing the empirical probability of heads
to be $1$ provides an entropy of $0$, which differs considerably from the true
entropy of a fair coin which is $1$ bit. The same effect occurs when one
computes the permission entropy conditioned on an irrelevant attribute
where only one observation per attribute value is available. In
\cite{Molloy:2010:MRN}, for instance, the last name of a user was found
to be highly relevant!
A practical solution is to compute $\rho_d(S)$ with only those values of
$S$ where sufficiently many observations are available. For instance, if
more than 10 users have the feature $s$=``Smith'', the empirical
probability $p(x_d|S=s)$ will give a good estimate of $h(x_d|S=s)$. In
our experiments, we neglected all attribute values with less than 10
observations.

\begin{figure}[htb]
\centering
\begin{tabular}{c}
\includegraphics[width=0.75\textwidth]{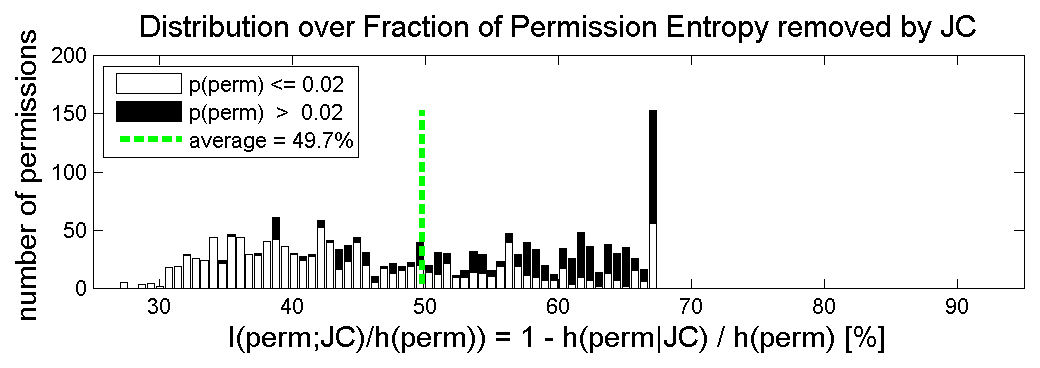} \\
\includegraphics[width=0.75\textwidth]{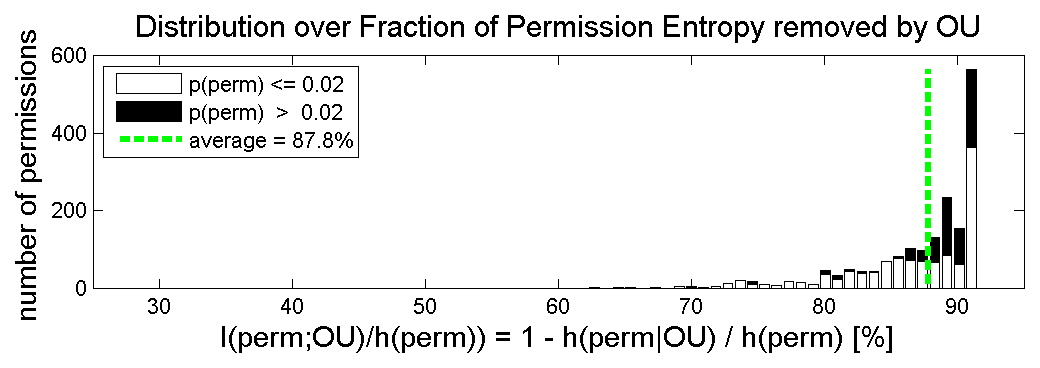}
\end{tabular}
\caption{\label{hybrid_fig:sIAnalysis2} Distribution of the measure of relevance 
Eq.~(\ref{hybrid_definition_rel}), the mutual information
weighted with inverse permission entropy.}
\end{figure}

We apply the proposed relevance measure to the two different user
attributes of the {\verb|LE-access|}
dataset, the job code (JC) and the organizational
unit (OU) of a user. The first attribute is
each user's so-called job code, which is a number that indexes the
kind of contract that the user has. We initially believed that this
attribute would be highly relevant for each user's permission as it is indicative
  of the user's business tasks.
We compute the relevance of these two attributes for each
permission. The results are depicted in
Figure~\ref{hybrid_fig:sIAnalysis2}. In these histograms, we count the
number of permissions for which the respective user attribute has the
given relevance score. As can be seen, the average reduction in
entropy is much higher for the OU (87.8\%) than for JC (49.7\%).  This
{result} means that knowledge of the OU almost determines most of the
permissions of the users while the JC provides relatively little information
about the permissions. We therefore only use OU in our role mining experiment.

{
\subsection{Results for hybrid role mining} 
}
\label{sec_HybRMresults}
We run experiments on the {\verb|LE-access|}
dataset. This time, we use the adapted E-step as derived in
Section~\ref{hybrid_sec:inference} with the organizational unit (OU) of the
user as the business attribute $S$. Again, we randomly split the data
into a training set and a test set, learn the roles on the training
set, and then compute the generalization error on the test set. This
time, we fix the number of roles to $k=30$ and only study the
influence of the business attributes on the result by varying the
mixing parameter $\lambda$.

To evaluate how well the resulting role configuration corresponds to
the business attributes of the user, we compute the average
conditional role entropy: 
\begin{equation}
h(\lset|S):= \! - \frac1N \sum_{i=1}^N
 \sum_{\lset}
p(z_{i\lset}|S=s_i)\log_2\left(p(z_{i\lset}|S=s_i)\right) \,.
\end{equation}
This number indicates how hard it is to guess the role assignments
$z_{i\lset}$ of a user $i$ given his organizational unit $s_i$. For a
good RBAC configuration this entropy is small, meaning that the knowledge of
the organizational unit suffices to determine the set of roles that the
user has. If the RBAC configuration does not correspond to the
organizational units at all, then knowledge of this attribute does not 
provide information about a user's roles and, as a result, the role 
entropy is high.
{The advantage of the conditional role entropy  over other measures
  of dispersion such as, for instance, nominal variance
  \cite{Domingo-FerrerS08} or attribute-spread \cite{italians2}, is
  that it directly resembles the task of an administrator who must
    assign roles to new users. When deciding which roles should be
      assigned to a user given his business attributes, there should be as little
        uncertainty as possible. For the same reason, we select relevant
          business information by relative mutual information instead of
  heuristic scores such as ``mineability'' \cite{newElicit}}.

\begin{figure}[htb]
\vspace{-0.5em}
  \centering
    \begin{minipage}[c]{0.7\linewidth}    
    \centering
\includegraphics[width=0.69\textwidth]  {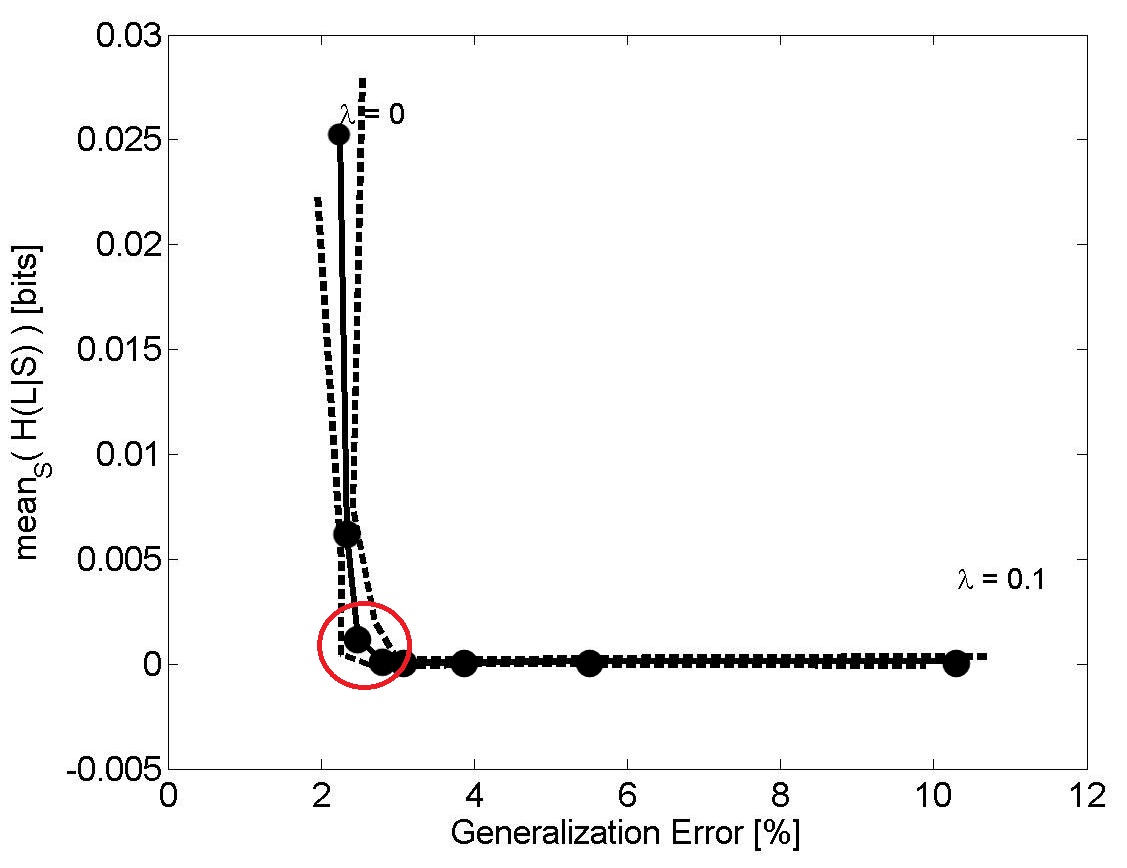}
    \end{minipage}
    \hfill
    \begin{minipage}[c]{0.29\linewidth}
      \caption{\label{hybrid_fig:HvsPred} Generalization error versus
  conditional role entropy for the experiment using organization
    units. The dashed lines represent the standard deviation over ten
  repeated experiments.}
    \end{minipage}
    \vspace{-0.5em}
\end{figure}

We depict the results in Figure~\ref{hybrid_fig:HvsPred}. We plot the
role entropy against the generalization error. Each point on this line 
is the median result over ten experiments with different random splits
in the training set and the test set. 
Each point is computed with a
different mixing parameter $\lambda$.  The plot demonstrates that by
increasing the influence of the organizational unit, the role entropy
decreases while the generalization error increases. This reduction
means that hybrid role mining with this attribute requires a
trade-off. Fortunately, most reduction in role entropy can be achieved
without significantly increasing the generalization error. When
further increasing $\lambda$, the generalization error increases
without significantly improving role entropy. This insensitivity
indicates a Pareto-optimum (marked with a circle in
Figure~\ref{hybrid_fig:HvsPred}), which defines the influence that one
should give to user attributes in hybrid role mining. Note that, we are able to find this point (to produce
Figure~\ref{hybrid_fig:HvsPred}) without knowledge of the true roles.

\section{Related work}
\label{relwork}
Shortly after the development of RBAC \cite{rbacOrig}, researchers and
practitioners alike recognized the importance of methods for role
engineering, see for instance \cite{Coyne:1996:RE:270152.270159}.  As
explained in the introduction, these methods can be classified as
being either top-down or bottom-up.  Top-down methods use process
descriptions, the organizational structure of the domain, or features
of the employees as given by the Human Resources Department of an
enterprise, to create roles.  \cite{Neumann}, for instance, present a
scenario-driven approach where a scenario's requirements are analyzed
according to its associated tasks. 
Permissions are then granted that
enable the task to be completed. 
In \cite{epstein}, a work flow is
proposed to manually engineer roles by analyzing business processes. Today, all pure top-down approaches are manually carried
out. Bottom-up role methods
have the advantage that they can be partially or fully automated by role mining algorithms, thereby relieving administrators from this
time-consuming and error-prone task.

 The first bottom-up role mining
algorithm was proposed in \cite{kuhlmann} and coined role mining.
Since then, a number of different bottom-up approaches have been
presented. For instance, the algorithm of \cite{orca} merges sets of
permissions to construct a tree of candidate roles.  Afterwards, it
selects the final roles from this tree such that, at every step, the 
permissions of a maximum number of users are covered. Subsequent role
mining algorithms usually followed a similar structure: {they} first 
construct a set of candidate roles and {afterwards} greedily select roles from
this set \cite{vaidya06roleminer,zhang,italians1}. These algorithms
differ from each other with respect to the proposal creation step or
the cost function used to select roles.  Vaidya et al.~formally
defined the role mining problem and variants in
\cite{JSVaidyaSacmat07rm} and \cite{Vaidya:2010:RMP:1805974.1805983}
and investigated the complexity of these problems. {The problem
  definitions proposed in these two papers} differ from our role
inference problem in that either the number of roles 
or the number of residuals tolerated when fitting the RBAC
configuration is given as input. Moreover, these definitions aim at a
high compression rate, while role inference aims at discovering {the} latent 
roles that underlie the data. 

The results presented in this paper build upon our prior work. In
\cite{RMdef}, we analyze the different definitions of role mining and
define the role inference problem.  In \cite{frank}, we derive DDM from
the deterministic permission assignment rule in RBAC. In
\cite{icml-paper}, we propose the MAC model, which we explore further
in \cite{MAC_jmlr}, adding different noise processes. 
\cite{mtc_ECML} focuses on determining the number of roles for role
mining in particular, and selecting the number of patterns
in unsupervised learning in general.  Finally, in \cite{hybridRM_ccs09}
we propose the hybrid role mining algorithm that we 
 revisit in Section~\ref{hybrid_sec:risk}.
 
 {{In this paper, we extend and generalize our prior work.}
   In particular, we draw connections between all these concepts and
   approaches.  We thereby generalize them within one consistent
   framework that covers i) the definition of role mining, ii) the
   approach to solve it, and iii) methods to evaluate solutions. While in
   \cite{RMdef} we motivate the role inference problem by 
   analyzing the real-world requirements of RBAC, here we take
   a more direct approach. We identify the input that is usually
   available in realistic scenarios and directly derive the role 
   inference problem from these assumptions. Moreover, we subsume the  models of our prior work, DDM and MAC, within one core model. In this
   way, we analyze the relationships between them and highlight the
   influence of noise processes, role hierarchies, and constraints on
   role assignments. This theoretical comparison of the models is
   supplemented by an experimental comparison of DDM, MAC, and other
   models and algorithms. In addition, we provide a sound measure of
   confidence for user permission assignments and investigate how
   conservatively the proposed methods modify noisy assignments. 
   We also show that DDM is structurally equivalent to a MAC
   model constrained to every user only having one role. Finally, our
   paper contains the first experimental comparison of DDM and MAC
   on both synthetic and real-world data.
}

\section{Conclusion} \label{concl-sec}
We put forth that, in contrast to conventional approaches, role mining
should be approached as a prediction problem, not as a compression
problem. We proposed an alternative, the role inference problem, with
the goal of finding the RBAC configuration that most likely underlies
a given access control matrix. This problem definition includes the
hybrid role mining scenario when additional information about the
users or the system is available.  To solve the role inference
problem, we derived a class of probabilistic models and analyzed
several variants. On real-world access control matrices, our
models demonstrate robust generalization ability while other methods
are rather fragile and their generalization ability depends on the 
particular data analyzed.

\subsection*{Acknowledgement}
We thank Andreas Streich for 
discussions and the collaboration on hybrid role mining. 

\bibliographystyle{plain}
\bibliography{probRM}

\medskip
\begin{appendix}

{
\section{Appendix}
\subsection{Marginalization}
\label{app_derivLikelih}

We convert the deterministic formula Eq.~(\ref{eq_bitGenerationDeterm}) into a probabilistic version by marginalizing out the latent variables $\mathbf{u}_{kd}$ from the joint distribution for $\mathbf{u}_{kd}$ and $x_{id}$.
The joint distribution is
\begin{equation}
p(x_{id}=0,\mathbf{u}_{\allInd \sep d} \vert \boldsymbol \beta_{\allInd \sep d}, \mathbf{z}_{i\sep \allInd}) = p(x_{id}=0\vert \mathbf{u}_{\allInd \sep d}, \mathbf{z}_{i\sep \allInd}) \prod_k p(u_{kd} \vert \beta_{kd}) \ .
\end{equation}
As a single bit $x_{id}$ depends on the outcome of a full bitvector $\mathbf{u}_{\allInd \sep d}$ of length $K$, one must marginalize over all realizations of such bit-vectors.

Let $\Omega$ be the set of all possible binary vectors $\mathbf{u}_{\allInd \sep d}$. Then the likelihood for $x_{id}=0$ is
\begin{align}
p(x_{id}&=0 \vert  \boldsymbol \beta_{\allInd \sep d}, \mathbf{z}_{i\sep \allInd})
 \\ 
 &= \sum_{\mathbf{u}_{\allInd \sep d}\in \Omega} p(x_{id}=0,\mathbf{u}_{\allInd \sep d} \vert \boldsymbol \beta_{\allInd \sep d}, \mathbf{z}_{i\sep \allInd}) 
   \\
 &= 
 \sum_{\mathbf{u}_{\allInd \sep d}\in \Omega} 
 p(x_{id}=0\vert \mathbf{u}_{\allInd \sep d}, \mathbf{z}_{i \sep \allInd}) \prod_k p(u_{kd} \vert \beta_{kd})
  \\
 &= 
 \sum_{\mathbf{u}_{\allInd \sep d}\in \Omega}
 \prod_k \left(1-u_{kd}\right)^{z_{ik}} \left(\beta_{kd}^{1-u_{kd}} (1-\beta_{kd})^{u_{kd}}\right)  
\label{eq_Subst}
\\
 &= \sum_{\mathbf{u}_{\allInd \sep d}\in \Omega}   
  \left(\prod_{k:z_{ik}=1} \left(1-u_{kd}\right)^{z_{ik}} \beta_{kd}^{1-u_{kd}} (1-\beta_{kd})^{u_{kd}}\right)
 \label{eq_marg_line3}
  \left(\prod_{k:z_{ik}=0} \beta_{kd}^{1-u_{kd}}
    (1-\beta_{kd})^{u_{kd}}\right) \, .
   \end{align}
   In the step from (\ref{eq_Subst}) to  (\ref{eq_marg_line3}) we substituted the individual probabilities with their definitions Eq.~(\ref{eq_pOfU}) and Eq.~(\ref{eq_bitGenerationDeterm}).
   In the last step, we separated the bit-vectors $\mathbf{u}_{\allInd \sep d}$ into the two cases where $z_{ik}=1$ and $z_{ik}=0$. The first case cancels all contributions of the sum where $z_{ik}=u_{kd}=1$ and for $z_{ik}=1, u_{kd}=0$ only the factor $\beta_{kd}$ remains. Therefore, it is convenient to introduce a modified set of bit-vectors 
   $\Omega' \!= \!\left\{\mathbf{u}_{\allInd \sep d}\in \Omega \;\vert
     u_{kd}=0\  \text{,  for all } k \text{ with } z_{ik}=1 \right\} \subset\Omega$,
    i.e., the entries of $u_{kd}$ that are relevant for object $i$ are fixed to $0$.
    The likelihood then takes the following compact form
   \begin{align}
   &p(x_{id}\!=\!0 \vert  \boldsymbol \beta_{\allInd \sep d}, \mathbf{z}_{i\sep \allInd}) 
 \! = \!\!\!
 \sum_{ \mathbf{u}_{\allInd \sep d}'\in \Omega'} \!\left\{\!\!
 \left(\prod_{k:z_{ik}=1}\!\! \! \beta_{kd}\!\!\right)\!
  \left(\prod_{k:z_{ik}=0}\!\!\! \beta_{kd}^{1-u_{kd}'}
    (1-\beta_{kd})^{u_{kd}'}\!\right)\!\!\right\}\! \, .
   \label{eq_marg_line4}    
   \end{align}

   The sum in Eq.~(\ref{eq_marg_line4}) has $\left|\Omega'\right|$
   terms. Let us pick a particular $k''$ with $z_{ik''}=0$. Half of the
   vectors in $\Omega'$ have $u_{k''j}=1$ and the other half have
   $u_{k''j}=0$, whereas the remaining bits $k\neq k''$ are equal in
   both halves. We can therefore factor out the terms where bitvectors
   differ only at position $k''$ in Eq.~(\ref{eq_marg_line4}). This
   reduces the number of terms in the sum by a factor of two, whereby
   the sum now ranges over the modified set $\Omega''\subset\Omega'$,
   where all bits are varied except for $k''$.

  In the following steps, we recursively factor out such terms. This successively interchanges the sum and the product in Eq.~(\ref{eq_marg_line4}) and makes it easy to see, in Eq.~(\ref{App_eq_marg_line6}),  that all contributions with $z_{ik}=0$ sum up to $1$.   
   \begin{align}
   &p(x_{id}=0 \vert  \beta_{\allInd \sep d}, \mathbf{z}_{i\sep \allInd}) \\
 &=  
  \beta_{k''j}^{z_{ik''}}
 \underbrace{\left( (1-\beta_{k''j})+\beta_{k''j}  \right)}_{=1\;\forall k''}
 \label{App_eq_marg_line5}
 \sum_{  \mathbf{u''}_{\allInd \sep d}\in \Omega''} \;
 \Biggl( \prod_{ { \begin{array}{cl} k:z_{ik}=1\\  k\neq k'' \end{array} }}
 \beta_{kd}^{z_{ik}} \Biggr)
   \Biggl(\prod_{{ \begin{array}{cl} k:z_{ik}=0\\  k\neq k'' \end{array} }} \beta_{kd}^{1-u''_{kd}} (1-\beta_{kd})^{u''_{kd}}\Biggr)
\\
 &= 
 \left(\prod_{k}  \beta_{kd}^{z_{ik}} \right) 
  \Biggl(\prod_{k:z_{ik}=0} \;
  \underbrace{\sum_{u'_{k\sep d}\in \left\{0,1\right\}}   \beta_{kd}^{1-u'_{kd}} (1-\beta_{kd})^{u'_{kd}}}_{=1\;\forall k} 
  \Biggr)
 \label{App_eq_marg_line6} 
 =\prod_{k}  \beta_{kd}^{z_{ik}} 
\end{align}

\subsection{Hierarchies}
\label{app_hierarchy}

We compute the probability that a user is not assigned to a permission
in a two-level role hierarchy given the probabilities of independent
role hierarchy parameters. We 
start from Eq.~(\ref{eq_2layerRule_short}).
\begin{align}
\llap{$p$}\left(\overline{x_{id}}\  {\vert\mathbf{z}_{i  \sep \allInd}^{+} ,\mathbf{y}_{ \allInd \sep d}^{+}, V^{+}} \right) 
&=
p\left(\overline{\bigvee_{ k=1}^K \left[z_{ik} \wedge
    \left(\bigvee_{l=1}^L v_{kl}\wedge y_{ld}\right) \right]} \
{\vert \mathbf{z}_{i  \sep \allInd}^{+},\mathbf{y}_{ \allInd \sep d}^{+}, V^{+}}\right)          
\begin{array}{clrr}
1\!\!\le\!\! k\!\!\le \!\!K\!\!\\
\!\!1\!\!\le\!\! l\!\!\le\!\! L
\end{array}
     \\ 
&=
     \prod_{k}p \biggl( \overline{z_{ik}\wedge\underbrace{ \biggl( \bigvee_{l}v_{kl}\wedge
     y_{ld} \biggr) }_{=:u_{kd}}} {\ \vert  \mathbf{z}_{i  \sep \allInd}^{+} ,\mathbf{y}_{ \allInd \sep d}^{+}, \mathbf{v}_{ k \sep \allInd }^{+} } \biggr) 
\label{eq_sumExcl1}
     \\ 
&=
     \prod_{k}\! \Bigl(\! 
     \underbrace{p\!\left(\overline{z_{ik}}\!\wedge\!
     u_{kd}{\vert  z_{i  \sep k}^{+} ,\mathbf{y}_{ \allInd \sep d}^{+}, \mathbf{v}_{ k \sep  \allInd }^{+} }\right) \! + \!p\!\left(\overline{z_{ik}}\!\wedge\!
     \overline{u_{kd}}{\vert  z_{i  \sep k}^{+} ,\mathbf{y}_{ \allInd \sep d}^{+}, \mathbf{v}_{ k \sep  \allInd }^{+} }\right)}_{=p\left(\overline{z_{ik}}{\vert z_{i  \sep k}^{+} }\right){=1-z_{i  \sep k}^{+} }} 
    \! + \!
     p\!\left(z_{ik}\!\wedge\!     \overline{u_{kd}} {\vert  z_{i  \sep k}^{+} ,\mathbf{y}_{ \allInd \sep d}^{+}, \mathbf{v}_{ k \sep  \allInd }^{+} }\right)\!
\Bigr) 
\label{eq_sumExcl2}
     \\ 
&=
    { \prod_{k}\left(1-z_{i  \sep k}^{+}+z_{i  \sep k}^{+}\  p(\overline{u_{kd}}{\vert\mathbf{y}_{ \allInd \sep d}^{+}, \mathbf{v}_{ k \sep \allInd }^{+} }) \right)  } \label{eq_2layerRule}
\end{align} 
Note that in the step from (\ref{eq_sumExcl1}) to (\ref{eq_sumExcl2}), the correct
probability is only obtained when summing over the probabilities of
exclusive events (in particular: $\overline{a\wedge b}= \overline{a}\vee \overline{b}$ but $p\left(\overline{a\wedge b}\right)\neq p\left(\overline{a}\right)+ p\left(\overline{b}\right)$). Given the generation $
u_{kd} = \bigvee_{l} v_{kl}\wedge y_{ld}. 
$, we have that
\begin{eqnarray}
p(\overline{u_{kd}}{\vert\mathbf{y}_{ \allInd \sep d}^{+}, \mathbf{v}_{ k \sep \allInd }^{+} })
&=&
  \prod_{l}p\left(\overline{v_{kl}\wedge y_{ld}}{\vert y_{ l \sep d}^{+},  v_{ k \sep l }^{+} }\right) 
=
  \prod_{l}\left[
  p(\overline{y_{ld}}{\vert y_{ l \sep d}^{+}})
+p(y_{ld}               {\vert y_{ l \sep d}^{+}})
   p(\overline{v_{kl}}{\vert v_{ k \sep l}^{+}})
\right] 
\, .
\end{eqnarray} 
Substituting this into Eq.~(\ref{eq_2layerRule}) yields
\begin{equation}  
  \llap{$p$}\left(\overline{x_{id}}\  {\vert\mathbf{z}_{i  \sep \allInd}^{+} ,\mathbf{y}_{ \allInd \sep d}^{+}, V^{+}} \right) 
 =
{ \prod_{k}
\left(
1-z_{i  \sep k}^{+}
 +z_{i  \sep k}^{+}\  
\prod_{l}
  \left(
 1- y_{ l \sep d}^{+})
+y_{ l \sep d}^{+} (1- v_{ k \sep l}^{+})  
\right) 
\right) } \, .
\end{equation}

\subsection{Gibbs Sampling Algorithm for DDM}
\label{app_eviDDM}
 
Here, we derive all necessary distributions for the Gibbs sampling algorithm for DDM in Section~\ref{sec_GibbsSampl}.
We start by collecting all distributions that define the model.
The data likelihood of the model is 
\begin{align}
\label{app_DDMlikelihood}
p\left(\mathbf{X}\vert\mathbf{Z},\mathbf{Y},\boldsymbol{\beta}\right)
&= 
\prod_{k,l}\left(1-\beta_{kl} \right)^{n^{(1)}_{kl}}
\beta_{kl}^{n^{(0)}_{kl}} 
 \, ,
\end{align} 
with the counters $
n^{(1)}_{kl}=\sum_{\genfrac{}{}{0pt}{}{i:z_{ik}=1,}{j:y_{lj}=1}} \eeins_{\left\{x_{ij}=1\right\}}
$ and $
n^{(0)}_{kl}=\sum_{\genfrac{}{}{0pt}{}{i:z_{ik}=1,}{j:y_{lj}=1}}
\eeins_{\left\{x_{ij}=0\right\}} \ .
%
$
The parameters $\beta_{kl}$ are random variables themselves. They follow the Beta distribution
\begin{align}
P_b\left(\beta_{kl}; \gamma,\gamma \right) 
&= 
\frac{\Gamma(2\gamma)}{2 \Gamma(\gamma)} \left(\beta_{kl} (1-\beta_{kl}) \right)^{\gamma-1}
= B(\gamma,\gamma)^{-1} \left(\beta_{kl} (1-\beta_{kl}) \right)^{\gamma-1} \ .
\end{align}
Here $B(.,.)$ is the beta function, also known as ``Euler integral of the first kind''.

In each step, one must sample new assignments of user $i'$ to roles from the distribution
\begin{eqnarray}
\label{samplingEq}
p\left(z_{i'k}\!=\!1\vert \mathbf{X},\mathbf{z}_{i\neq i'\sep\allInd},\mathbf{Y}\right) 
&=&
\text{const} \cdot
 p\left(\mathbf{X}\vert \mathbf{Z},\mathbf{Y}\right)
p\left(z_{i'k}\!=\!1\vert\mathbf{z}_{i\neq i'\sep\allInd}\right) \ .
\end{eqnarray}
Assigning permissions to roles (updating $\mathbf{Y}$)
has the same form, just with $\mathbf{Y}$ and $\mathbf{Z}$ interchanged.
In order to compute this term for a particular user $i'$, one must compute the evidence term $p\left(\mathbf{X}\vert \mathbf{Z},\mathbf{Y}\right)$ and the Dirichlet process prior $p\left(z_{i'k}\!=\!1\vert\mathbf{z}_{i\neq i'\sep\allInd}\right)$ for all available roles $k\in\left\{1,\myldots,K\right\}$ and for a potential new role with index $K+1$. 
We introduced the Dirichlet process prior 
 $p\left(z_{i'k}\!=\!1\vert\mathbf{z}_{i\neq i'\sep\allInd}\right)$ in Section~\ref{eq_PriorAssumption}.
The evidence term is
\begin{align}
p\left(\mathbf{X}\vert\mathbf{Z},\mathbf{Y}\right) 
&= \int  
p\left(\mathbf{X},\boldsymbol{\beta}\vert \mathbf{Z},\mathbf{Y}\right)
 d\boldsymbol \beta
 = \frac{p\left(\mathbf{X}\vert
     \mathbf{Z},\mathbf{Y},\boldsymbol{\beta}\right)
   p\left(\boldsymbol{\beta}\vert
     \mathbf{Z},\mathbf{Y}\right)}{p\left(\boldsymbol{\beta}\vert
     \mathbf{Z},\mathbf{Y}, \mathbf{X}\right)}  \, .
 \label{app_DDMevidence}
\end{align}
So far, we just applied Bayes' rule. The numerator
is a product of terms that we already have: the data likelihood in Eq.~(\ref{app_DDMlikelihood}) and the Beta distributions
$p\left(\boldsymbol{\beta}\vert \mathbf{Z},\mathbf{Y}\right) = \prod_{k,l} P_b\left(\beta_{kl}; \gamma,\gamma \right)$. 
The term in the denominator is the posterior probability
$p\left(\boldsymbol{\beta}\vert \mathbf{Z},\mathbf{Y},
  \mathbf{X}\right)$ (the probability of $\boldsymbol{\beta}$
\emph{after} having observed the data $\mathbf{X}$). We can rewrite this
(again using Bayes) to
\begin{align}
p\left(\boldsymbol{\beta}\vert \mathbf{Z},\mathbf{Y}, \mathbf{X}\right)
&=   \frac{p\left(\mathbf{X}\vert\mathbf{Z},\mathbf{Y},\boldsymbol{\beta}\right)   \prod_{k,l} P_b\left(\beta_{kl}; \gamma,\gamma \right)}{p\left(\mathbf{X}\vert\mathbf{Z},\mathbf{Y}\right)  }
\\
&= \text{const} \cdot  p\left(\mathbf{X}\vert\mathbf{Z},\mathbf{Y},\boldsymbol{\beta}\right)   \prod_{k,l} P_b\left(\beta_{kl}; \gamma,\gamma \right)
\\
&= \text{const}\cdot
\prod_{k,l}\left(1-\beta_{kl} \right)^{n^{(1)}_{kl}+\gamma-1}
\beta_{kl}^{n^{(0)}_{kl}+\gamma-1} \, .
\end{align}
 Comparison with the Beta distribution makes it easy to identify how this probability must be normalized. Therefore, we can analytically compute the posterior:
\begin{align}
p\left(\boldsymbol{\beta}\vert \mathbf{Z},\mathbf{Y}, \mathbf{X}\right)
&=
\prod_{k,l} \frac{\Gamma(n^{(1)}_{kl}+n^{(0)}_{kl}+2\gamma)}{ \Gamma(n^{(1)}_{kl}+\gamma) \Gamma(n^{(0)}_{kl}+\gamma)} 
\left(1-\beta_{kl} \right)^{n^{(1)}_{kl}+\gamma-1}
\beta_{kl}^{n^{(0)}_{kl}+\gamma-1} 
\\
&= 
\prod_{k,l} B(n^{(1)}_{kl}+\gamma,n^{(0)}_{kl}+\gamma)^{-1}  
\left(1-\beta_{kl} \right)^{n^{(1)}_{kl}+\gamma-1}
\beta_{kl}^{n^{(0)}_{kl}+\gamma-1} \ .
\end{align}
The Beta distribution is a conjugate prior of the Bernoulli
distribution. As a consequence, the posterior of a Bernoulli likelihood
and a Beta prior again has the form of a Bernoulli distribution as we just observed in the last derivation.
Substituting the posterior back to Eq.~(\ref{app_DDMevidence}) results in the analytic expression of the evidence term.
\begin{align}
p\left(\mathbf{X}\vert\mathbf{Z},\mathbf{Y}\right) 
&= \prod_{k,l}
\frac{B(n^{(1)}_{kl}+\gamma,n^{(0)}_{kl}+\gamma)}{B(\gamma,\gamma)}
\end{align}
Computation of this term only involves updating the two counters.

\subsection{Marginalization of the noise indicator $\xi$}
\label{app_margXi}
Let us assume the random noise indicator $\xi_{i\sep d}$ follows a Bernoulli distribution 
\begin{equation}
p\left(\xi_{i\sep d}\left.\right|\epsilon\right)
    = \epsilon^{\xi_{i\sep d}}\left(1-\epsilon\right)^{1-\xi_{i\sep d}}
     \ .
\end{equation}
The structural bit is generated by the structure model $p_{\flatRBAC}\left(x_{i\sep d} \left.\right|            \mathbf{Z},\boldsymbol{\beta} \right)       $ (Eq.~\ref{eq_flatRBAC}), the noise bit is generated by the Bernoulli distribution 
$p_N \left(x^N_{i\sep d}\left.\right|r\right)$,
 and the observable bit is generated by
\begin{equation}
  x_{i\sep d} = (1-\xi_{i\sep d}) x^S_{i\sep d} + \xi_{i\sep d} x^N_{i\sep d} \  .
\end{equation}
Using these distributions, the joint probability of $x_{i\sep d}$ and $\xi_{i\sep
d}$ is
\begin{equation}
p\left(x_{i\sep d}, \mathbf{\xi}\left.\right|
        \mathbf{Z},\boldsymbol{\beta}, r, \epsilon \right)
  =
  \prod_{i\sep d}
        \left(\epsilon \cdot p_N \left(x_{i\sep d}\left.\right|r\right)\right)^{\xi_{i\sep d}}\cdot
        \left((1-\epsilon) \cdot p_{\flatRBAC}\left(x_{i\sep d}
            \left.\right|            \mathbf{Z},\boldsymbol{\beta}
          \right)\right)^{1-\xi_{i\sep d}} \, .
 \end{equation}            
We sum over all possible outcomes of all noise indicators and obtain the final likelihood of this mixture noise model.
\begin{align}
 p_M\left(\mathbf{X}\left.\right|\mathbf{Z},\boldsymbol{\beta}, r,
        \epsilon \right)
&= \sum_{\left\{\mathbf{\xi}\right\}} \prod_{i\sep d}
       \left(\epsilon \cdot p_N \left(x_{i\sep d}\left.\right|r\right)\right)^{\xi_{i\sep d}}\cdot
        \left((1-\epsilon) \cdot p_{\flatRBAC}\left(x_{i\sep d} \left.\right|            \mathbf{Z},\boldsymbol{\beta} \right)\right)^{1-\xi_{i\sep d}} \\
&= \prod_{i\sep d} \sum_{\xi_{i\sep d}\!\in\!\{0,1\}}
       \left(\epsilon \cdot p_N \left(x_{i\sep d}\left.\right|r\right)\right)^{\xi_{i\sep d}}\cdot
        \left((1-\epsilon) \cdot p_{\flatRBAC}\left(x_{i\sep d} \left.\right|            \mathbf{Z},\boldsymbol{\beta} \right)\right)^{1-\xi_{i\sep d}}
 \\
 &= \prod_{i\sep d} \left(
        \epsilon \cdot p_N \left(x_{i\sep d}\left.\right|r\right)
        + (1-\epsilon) \cdot p_{\flatRBAC}\left(x_{i\sep d} \left.\right|            \mathbf{Z},\boldsymbol{\beta} \right) \right)
 \ . 
\end{align}

\subsection{Update equations for deterministic annealing}
\label{ssec:Updates_generalForm}

We define the empirical risk of assigning a
user $i$ to the set of roles $\lset$ as the negative
log-likelihood of the permissions $x_{i\sep \allInd}$ being generated by
 role set $\lset$, given parameters $\Theta= (\beta,\epsilon,r)$:
\begin{eqnarray}
R_{i\sep \lset}
    &:=& - \log p(x_{i\cdot}| \lset, \Theta) \nonumber
= - \sum_d \log \left(  x_{id} \left( 1-q_{\lset\sep d} \right)
        +(1-x_{id})q_{\lset\sep d} \right)
    \label{eq:risk_generalForm} \ ,
\end{eqnarray}
where for the auxiliary variable we have that
$
q_{\lset \sep d}
    :=  \epsilon r + (1-\epsilon)\left(1-\beta_{\lset\sep d}\right) 
$.

The \textbf{responsibility} $\gamma_{i\lset}$ of the
assignment-set $\lset$ for data item $i$ is given by
\begin{eqnarray}
\gamma_{i\lset}
    := \frac{\exp\left( -R_{i\sep \lset} /T\right)}
        {\sum_{\lset'\in\lsetset}
            \exp\left( -R_{i\sep \lset'} / T\right)}\ .
\end{eqnarray}
In this way, the matrix $\boldsymbol{\gamma}$ defines a
probability distribution over the space of all clustering solutions.
The expected empirical risk $\expectz{G}{R}$ of the solutions
under this distribution $G$ is $
 \expectz{G}{R_{i\sep \lset}}
    = \sum_i \sum_{\lset}
    \gamma_{i\sep \lset} R_{i\sep \lset}    
$. Finally, the \textbf{partition function} $Z$ and the \textbf{free energy}
$F$ are defined as follows.
\begin{align}
Z
    := \prod_i \sum_{\lset} \exp\left( - R_{i\sep \lset} /T \right)
\quad \quad F
    := -T \log Z
    =-T \sum_i \log \left( \sum_{\lset}
        \exp\left( - R_{i\sep \lset} / T\right) \right)  \, .
\end{align}

Given the above, we derive the updates of the model parameters
based on the first-order condition of the free energy $F$.
We therefore introduce the generic model parameter $\theta$, which
stands for any of the model parameters, i.e.~$\theta \in
\left\{\beta_{\mu\nu},\epsilon, r\right\}$.
Here, $\mu$ is some particular value of role index $k$ and $\nu$ is some
particular value of permission index $d$. Using this notation, the derivative of the free
energy with respect to $\theta$ is given by
\begin{align}
\frac{\partial F}{\partial \theta}
    =   \sum_i  \sum_{\lset}
         \gamma_{i\sep \lset} \frac{\partial R_{i\sep \lset}}
                {\partial \theta}
                =  
                 \sum_i  \sum_{\lset}
         \gamma_{i\sep \lset} \sum_d
            \frac{(1-2x_{id})\frac{\partial q_{\lset\sep d}}
            {\partial \theta}}{x_{id} \left( 1-q_{\lset\sep d} \right)
            +(1-x_{id})q_{\lset\sep d} } \ .
\end{align}

The partial derivatives with respect to
$\theta\in\left\{\beta_{\mu\nu},\epsilon, r\right\}$ are:
\begin{align}
\frac{\partial }{\partial \beta_{\mu\nu}} q_{\lset\sep d}
    = \left(1-\epsilon\right) \beta_{\lset \setminus  \{\mu\}, d}\;
        \eeins_{\{\nu=d\}} \eeins_{\{\mu\in\lset\}} \ ,
        \ \ \ \
\frac{\partial }{\partial \epsilon} q_{\lset\sep d}
    = 1-r - \beta_{\lset\sep  d} \ ,
    \ \ \ \
\frac{\partial }{\partial r} q_{\lset\sep d}
    = -\epsilon \ .
\end{align}
This results in the following first-order conditions for the individual role parameter updates and the noise parameter updates:
\begin{align}      
  \beta_{\mu\nu}&: \      
     \sum_{\lset: \mu\in \lset} \beta_{\lset\setminus  \{\mu\}, \nu}
     \left\{ \frac{
             \sum_{i:x_{i \nu}=1} \gamma_{i\sep \lset}}
            {\epsilon r + (1-\epsilon) \left(1-\beta_{\lset\sep \nu}\right)}
        -\frac{  \sum_{i:x_{i \nu}=0}  \gamma_{i\sep \lset}}
             {1-\epsilon r - (1-\epsilon) \left(1-\beta_{\lset\sep \nu}\right)}
             \right\}= 0
             \label{eq_betaUpdate}
\\
     \epsilon&:\ 
   \sum_{d} \left\{\! \sum_{\lset}
        \frac{(1\!-\!r\!-\!\beta_{\lset\sep d})\sum_{i: x_{id}\!=\!1}\gamma_{i\sep \lset} }
            { \epsilon r + (1-\epsilon) \left(1-\beta_{\lset\sep d}\right)}
            - \sum_{\lset}
        \frac{(1\!-\!r\!-\!\beta_{\lset\sep d})\sum_{i: x_{id}\!=\!0}\gamma_{i\sep \lset}  }
             {  1-\epsilon r - (1-\epsilon) \left(1-\beta_{\lset\sep d}\right)}\!\right\} = 0 
              \label{eq_epsUpdate}     
              \\
             \label{eq_rUpdate}
    r&:\  
   \sum_{d} \left\{ \sum_{\lset}
        \frac{\sum_{i: x_{id}=0}\gamma_{i\sep \lset}  }
            {  1-\epsilon r - (1-\epsilon) \left(1-\beta_{\lset\sep d}\right)}
            - \sum_{\lset}
        \frac{\sum_{i: x_{id}=1}\gamma_{i\sep \lset} }
            { \epsilon r + (1-\epsilon) \left(1-\beta_{\lset\sep d}\right)} \right\}= 0
\end{align}
Here, $\beta_{\lset\setminus  \{\mu\}, \nu}=\prod_{k\in\lset,k\neq\mu} \beta_{k\nu}$ and $\sum_{\lset:\mu\in \lset}$ is the sum over all role sets containing role $\mu$.
As there is no analytic expression for the solutions of the above equations, we use Newton's method we find the root.

}

\end{appendix}

\end{document}